\documentclass[useAMS,usenatbib]{mn2e}

\usepackage{framed}
\usepackage{algpseudocode}
\usepackage{amssymb}
\usepackage{epsfig}
\usepackage{graphicx}
\usepackage{epstopdf}

\title[Modelling the ACPS of a Solar-Type Oscillator]{Modelling the Autocovariance of the Power Spectrum of a Solar-Type Oscillator}
\author[T. L. Campante et al.]{T. L. Campante$^{1,2}$\thanks{E-mail:
campante@astro.up.pt; campante@phys.au.dk}, C.
Karoff$^{2,3}$, W. J. Chaplin$^{3}$, Y. P. Elsworth$^{3}$, R. Handberg$^{2}$ \newauthor and S. Hekker$^{3}$\\
$^{1}$Centro de Astrof\'isica, Faculdade de Ci\^encias, Universidade do Porto, Rua das Estrelas, 4150-762 Porto, Portugal\\
$^{2}$Danish AsteroSeismology Centre, Department of Physics and Astronomy, University of Aarhus, 8000 Aarhus C, Denmark\\
$^{3}$School of Physics and Astronomy, University of Birmingham, Edgbaston, Birmingham B15 2TT, UK}

\begin{document}



\maketitle

\label{firstpage}

\begin{abstract}
Asteroseismology is able to conduct studies on the interiors of solar-type stars from the analysis of stellar acoustic spectra. However, such an analysis process often has to rely upon subjective choices made throughout. A recurring problem is to determine whether a signal in the acoustic spectrum originates from a radial or a dipolar oscillation mode. In order to overcome this problem, we present a procedure for modelling and fitting the autocovariance of the power spectrum which can be used to obtain global seismic parameters of solar-type stars, doing so in an automated fashion without the need to make subjective choices. From the set of retrievable global seismic parameters we emphasize the mean small frequency separation and, depending on the intrinsic characteristics of the power spectrum, the mean rotational frequency splitting. Since this procedure is automated, it can serve as a useful tool in the analysis of the more than one thousand solar-type stars expected to be observed as part of the \emph{Kepler} Asteroseismic Investigation (KAI). We apply the aforementioned procedure to simulations of the Sun. Assuming different apparent magnitudes, we address the issues of how accurately and how precisely we can retrieve the several global seismic parameters were the Sun to be observed as part of the KAI.
\end{abstract}


\section{Introduction}
Seismology of solar-type oscillators is a powerful tool that can be used to increase our understanding of stellar structure and evolution. Oscillations in main-sequence stars and subgiants have been measured thanks to data collected from ground-based high-precision spectroscopy \citep[for a review see e.g.,][]{BeddingKjeldsen} and, more recently, to photometric space-based missions such as \emph{CoRoT} \citep[see e.g.,][]{Appourchaux,Michel}. The \emph{Kepler} mission  \citep[for a discussion on the expected results of the asteroseismic investigation see][]{Kepler,Karoff} will lead to a revolution in the field of asteroseismology of solar-type oscillators, since it will increase by more than two orders of magnitude the number of stars for which high-quality observations will be available, while allowing for long-term follow-ups of a selection of these targets. As of the time of writing of this article, first results arising from the \emph{Kepler} asteroseismic programme had already been made available \citep[][]{Kepler6,Kepler3,Kepler1,Kepler2,Kepler5,Kepler4,Kepler7}.   

Due to the large number of stars observed with \emph{Kepler}, automated and innovative analysis pipelines/tools are needed in order to cope with the plenitude of available data \citep[see e.g.,][]{Hekker,Huber,Kallinger,Karoff10,Mathur,Mosser09,Roxburgh}. The automated pipelines dedicated to the analysis of acoustic spectra that have been developed so far aim mainly at measuring the frequency of maximum amplitude, $\nu_{\rm max}$, the maximum mode amplitude, $A_{\rm max}$, and the mean large frequency separation, $\Delta \nu$. In the present study we give continuity to this work by presenting a tool capable of modelling and fitting the ACPS{\footnote{To be precise, we compute the autocovariance of the power density spectrum.}} (AutoCovariance of the Power Spectrum) of a solar-type oscillator. The current version of the tool accepts as free model parameters the mean small frequency separation parameter, $D_0$ (see \S\ref{sec:mfreq} for the definition of this parameter), the mean rotational frequency splitting, $\nu_{\rm{s}}$, the mean linewidth, $\langle\Gamma\rangle$, and the stellar inclination angle, $i$. The output generated by this tool will hopefully contribute to further explore the diagnostic potential of solar-like oscillations, especially that of the large and small separations \citep[see e.g.,][]{Dalsgaard}. These two quantities can in turn be used to provide estimates of the radii, masses and ages of solar-like stars with consistent uncertainties \citep[][]{Karoff10}.

We start in \S\ref{sec:rationale} by providing a logical basis that led us to the development of this particular analysis tool. This is followed by a thorough description of its implementation in \S\ref{sec:implementation}. In \S\ref{sec:results} we apply the tool to simulated data, more specifically, to a solar analog as it would be seen by \emph{Kepler}. A discussion and conclusions are presented in \S\ref{sec:conclusions}.
 
\section{Rationale Behind the ACPS-Modelling Procedure}\label{sec:rationale}
Unlike a canonical \emph{peak-bagging} procedure \citep*[see e.g.,][]{Anderson}, in the framework of the ACPS-modelling procedure individual modes do not need to be tagged by angular degree, i.e.~by wave number $\ell$. This is required in peak-bagging to ensure the correct model is fitted to the observed modes. Instead we build a model that describes both the global and average properties of the most prominent modes, which are in turn encoded on a small set of free model parameters. This might be regarded as a major advantage, especially if we recall the ambiguity found in mode (angular degree) identification concerning the \emph{CoRoT} target HD 49933, only recently solved after a new longer time series was made available \citep[][]{Appourchaux,Benomar2}. 

As it is currently implemented, the ACPS-modelling procedure assumes the small frequency separation to be constant and not a function of frequency. Especially for evolved stars this was thought not to be a good approximation \citep{Soriano}. However, in the light of the first \emph{Kepler} results on red giants \citep[][]{Kepler6}, it is perfectly valid to assume -- at least for low-luminosity red giants -- a mean value of the small frequency separation between adjacent modes with $\ell\!=\!0$ and $\ell\!=\!2$, $\delta\nu_{02}$, it being in fact a nearly constant fraction of $\Delta \nu$. Therefore, the procedure presented herein can in principle also be employed in the case of evolved stars displaying solar-like oscillations. The ACPS-modelling procedure will however not provide sensible output when mixed modes \citep*[see e.g.,][]{Aizenman} are present. In any event, the presence of mixed modes will manifest clearly in the ACPS, either as a significant broadening of the peaks or by the appearance of extra peaks.

\begin{figure}
\centering
\begin{tabular}{c}
\epsfig{file=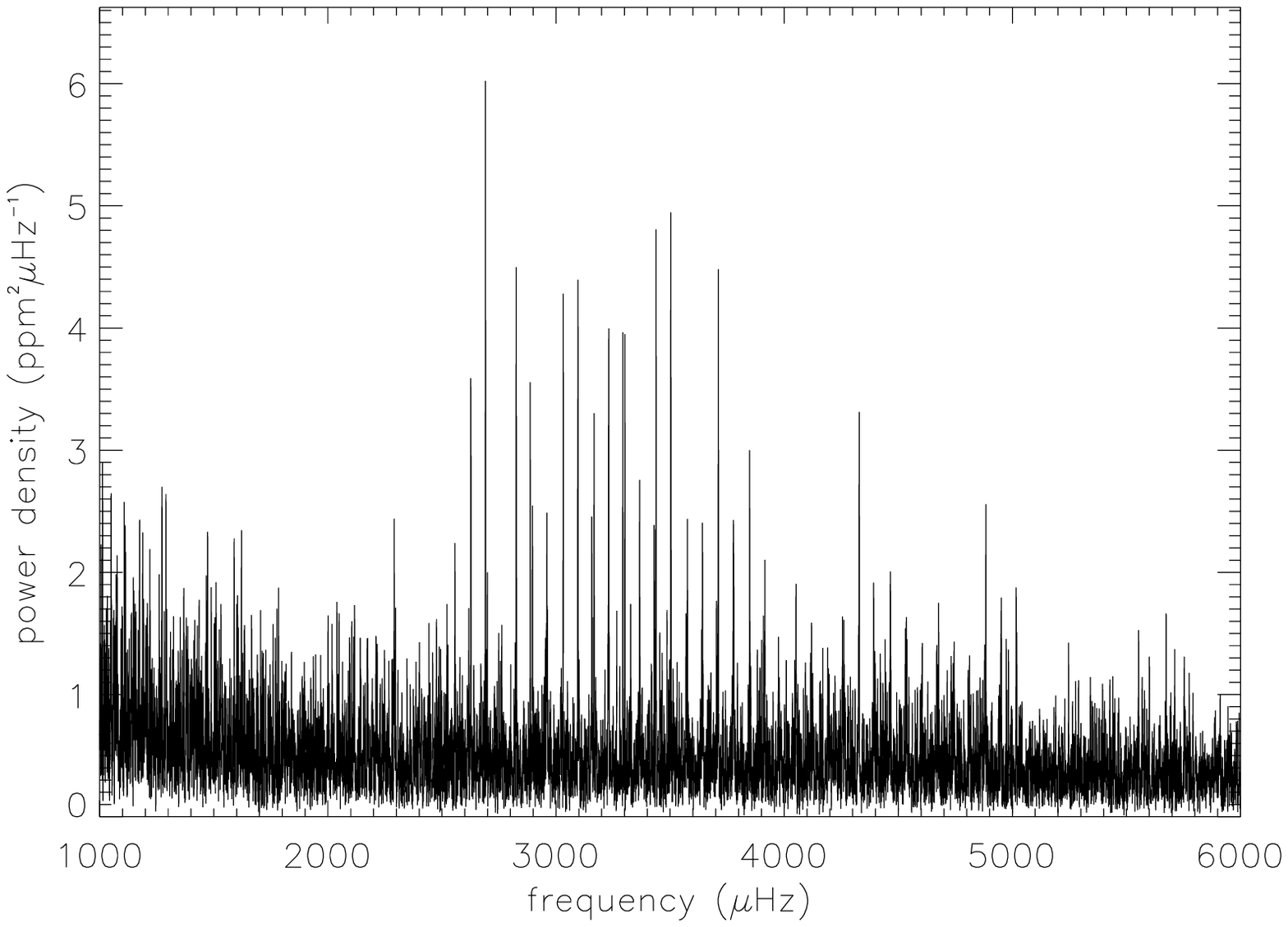,width=0.9\linewidth,clip=} \\ 
\epsfig{file=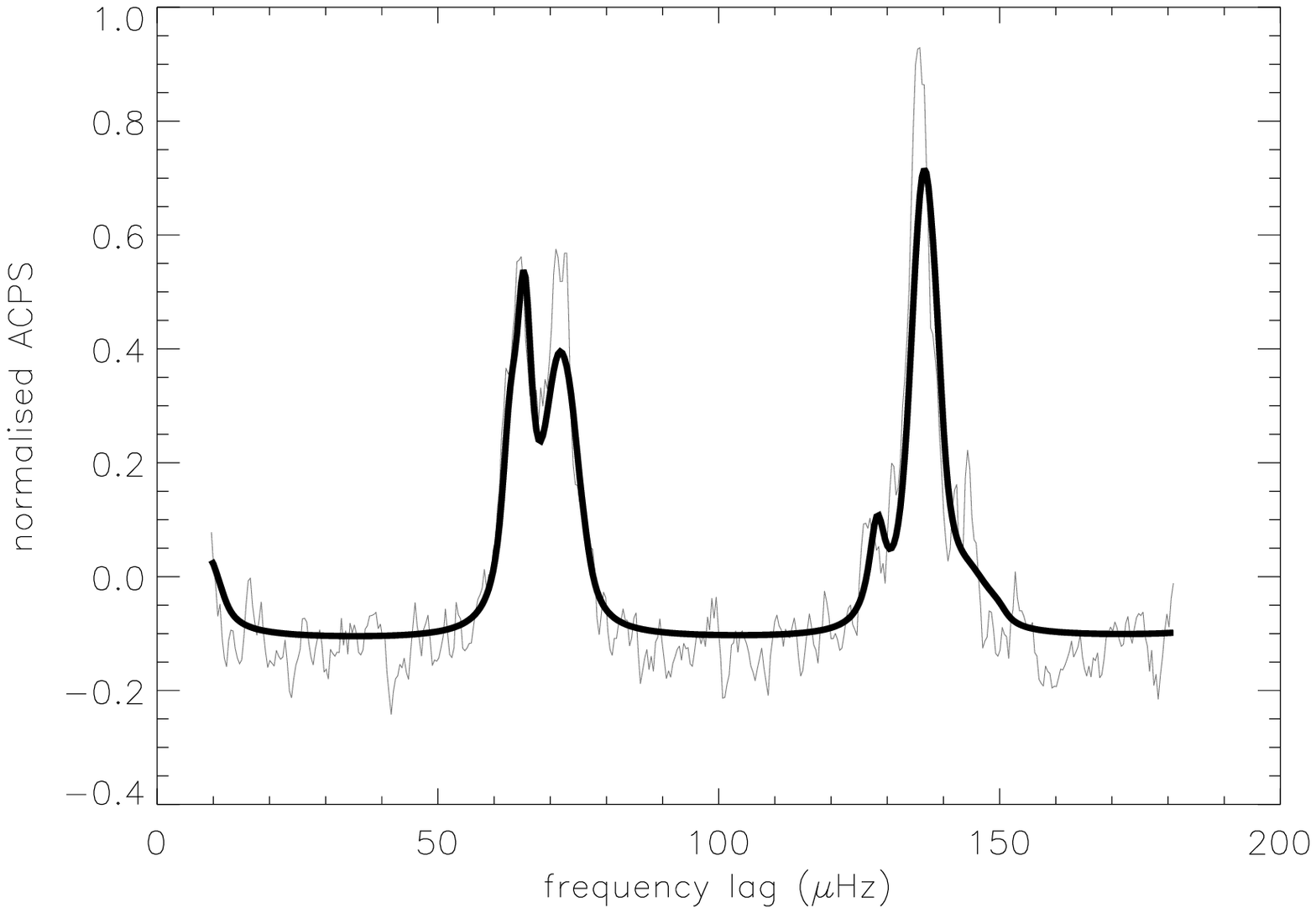,width=0.9\linewidth,clip=}
\end{tabular}
\caption{\emph{Top Panel:} Artificial acoustic spectrum of a solar analog evidencing a regular frequency structure. \emph{Bottom Panel:} Fit (black solid line) to the corresponding ACPS obtained using the MAP as our choice of summary statistic to represent the parameter posterior distributions. In order to enhance the visibility of the features in the ACPS we have slightly smoothed it by running a boxcar average (this is done both to the observed and to the model ACPS). Notice that the artificial star (Boris) has $\Delta\nu\!\simeq\!135 \: {\rm{\mu Hz}}$ (see discussion in \S\ref{sec:results}).}
\label{fig_Diagnose4}
\end{figure}

The way information is presented in the ACPS makes it very amenable for fitting. In fact, the ACPS of a solar-type oscillator will show prominent features at multiples of half the large frequency separation, a clear manifestation of the regular frequency structure of the acoustic spectrum. The shape of these features will depend upon the $\delta\nu_{02}$ and $\delta\nu_{01}$ spacings (see \S\ref{sec:mfreq} for a definition of the latter), mode lifetime, rotation and stellar inclination. Fig.~\ref{fig_Diagnose4} displays an artificial acoustic spectrum of a solar analog (top panel) as it would be obtained from a 30-day long \emph{Kepler}'s time series together with a fit to its ACPS (bottom panel). Notice the prominent features in the ACPS around $135\:{\rm{\mu Hz}}$ and half that value. The latter feature is clearly split into two peaks which constitutes a signature of $\delta\nu_{01}$. A signature of the $\delta\nu_{02}$ spacing is also present in the wings of the feature at $\Delta \nu$. The widths of either feature are generally a signature of the mode lifetime of the oscillations, whereas the presence of any fine structure is a signature of rotation and stellar inclination. The ACPS depicted in Fig.~\ref{fig_Diagnose4} does indeed provide us with information about the large and small frequency separations, and mode lifetime. The question we address first is thus: \emph{How do we extract this information from the autocovariance of the power spectrum?}

\section{The ACPS-Modelling Procedure}\label{sec:implementation}
We intend to fit the observed ACPS to an ACPS created from a model of the p-mode power density spectrum (PDS) that is described by a slightly modified asymptotic relation, which in turn depends only on a few free parameters (\S\ref{sec:modelling}). Although we do not fully explore its potential, we opt for a Bayesian approach to the fitting problem for the simple reason that any prior information might be taken into account, particularly useful when constraining the parameter space (\S\ref{sec:bayesian}). Most importantly, we aim at obtaining a full joint posterior probability density function (PDF) concerning the set of model parameters and hence we end up employing a Markov chain Monte Carlo (MCMC) sampler (\S\ref{sec:MCMC}).

\subsection{Modelling the observed ACPS}\label{sec:modelling}
The ACPS is computed as a function of the frequency lag, $L$, according to the following expression:
\begin{equation}
\label{eq_Autocovariance}
	{\rm{ACPS}}(L)=\frac{1}{N} \sum_{j=0}^{N-L-1} (P_j-\overline{P}) \, (P_{j+L}-\overline{P}) \, ,
\end{equation}
where $P_j$ is the particular value taken by the PDS at the fixed frequency bin $j$, $\overline{P}$ is the mean value of the sample population $\{P_j\}$, i.e.~the mean value of the PDS within the frequency interval of interest, and $N$ is the total number of bins within this same frequency interval. The frequency interval of interest should be in accordance with the p-mode range. The lag, $L$, which is expressed in terms of number of frequency bins, is allowed to vary between the corresponding frequency values of $10\:{\rm{\mu Hz}}$ and $4/3 \, \Delta\nu$. The reason why we constrain $L$ to this range has to do with the fact that we are primarily interested in measuring the small frequency separation. The signature of the small frequency separation is in fact mainly conveyed by the features at $\Delta\nu/2$ and $\Delta\nu$. For $L$ less than $10\:{\rm{\mu Hz}}$ the ACPS is dominated by the signature of the window function, whereas for $L$ greater than approximately $4/3 \, \Delta\nu$ it reproduces the features seen below this value of the lag and thus becoming redundant. Finally, the ACPS is normalised after division by its maximum value.

As already stated, modelling the observed ACPS consists of two steps: (i) building a model of the subjacent p-mode PDS and (ii) computing its autocovariance as explained above. In what follows we provide the reader with the basic theory behind the power spectrum of solar-like oscillations and end up describing how the model p-mode spectrum is generated.

\subsubsection{The model}
Solar-like oscillations are global standing acoustic waves, also known as p modes, driven by near-surface turbulent convection \citep[see e.g.,][]{Balmforth2,Balmforth1}. Radial oscillation modes are characterised by the radial order $n$, whereas non-radial modes are additionally characterised by the non-radial wave numbers $\ell$ and $m$. 

Ignoring any departure from spherical symmetry, non-radial modes differing only on the azimuthal wave number $m$ are degenerate. Stellar rotation removes the $(2\ell + 1)$-fold degeneracy of the frequency of oscillation of non-radial modes. When the angular velocity of the star, $\Omega$, is small and in the case of rigid-body rotation, the frequency of a $(n,\ell,m)$ mode is given to first order by \citep[][]{Ledoux1951}:
\begin{equation}
\label{eq_Splitting}
	\nu_{n\ell m} = \nu_{n\ell} + m\frac{\Omega}{2\pi}(1-C_{n \ell}) \, .
\end{equation}
The kinematic splitting, $m\Omega/(2\pi)$, is corrected for the effect of the Coriolis force through the dimensionless quantity $C_{n\ell}\!>\!0$. In the asymptotic regime, i.e.~for high-order, low-degree p modes, rotational splitting is dominated by advection and the splitting between adjacent modes within a multiplet is $\nu_{\rm{s}}\!\simeq\!\Omega/(2\pi)$. Typically, $\nu_{{\rm{s}}\odot}\!=\!0.4 \: {\rm{\mu Hz}}$. 

The PDS of a single mode of oscillation is distributed around a limit spectrum with an exponential probability distribution \citep[][]{Woodard1984,DuvallHarvey}. This limit spectrum contains the information on the physics of the mode and may be described as a standard Lorentzian profile near the resonance. It follows that the overall limit p-mode spectrum is given by \citep[see e.g.,][]{FletcherEtAl.2006}:
\begin{eqnarray}
		 \mathcal{P}(\nu; S_{n \ell m}, \nu_{n\ell}, \nu_{\rm{s}}, \Gamma_{n\ell m}, N_{\nu})  = \nonumber \\
		 \sum_{n=n_0}^{n_{\rm{max}}} \sum_{\ell=0}^{\ell_{\rm{max}}} \sum_{m=-\ell}^\ell \,  \frac{S_{n\ell m}}{ 1 + \left(\frac{2 \, (\nu-\nu_{n\ell}-m\nu_{\rm{s}})}{\Gamma_{n\ell m}}\right)^2 } + N_{\nu}(\nu) \, ,
\label{eq_LimitSpectrum}
\end{eqnarray}
where $S_{n \ell m}$ is the mode height, $\nu_{n\ell}$ is the central frequency of the $(n,\ell)$ multiplet and $\Gamma_{n\ell m}$ is the mode linewidth. In the case of solar-type oscillators and for low angular degree $\ell$, we can assume that $\Gamma$ is a function of frequency alone. $\Gamma$ is related to the mode lifetime, $\tau$, through $\Gamma\!=\!(\pi\tau)^{-1}$. $N_{\nu}$ describes the background signal originating from both granulation and activity. Notice that we are assuming that a mode is uncorrelated with any other modes or with the background signal. The stellar background signal is commonly modelled as a sum of power laws describing these physical phenomena \citep*[][]{Harvey,AigrainEtAl.2004}: 
\begin{equation}
\label{eq_Background}
	N_{\nu}(\nu) = \sum_{k=1}^{k_{\rm{max}}}\frac{A_k^2 B_k}{1+(2\pi B_k\nu)^{C_k}} + N \, ,
\end{equation}
$\{A_k\}$ and $\{B_k\}$ being, respectively, the corresponding amplitudes and characteristic time scales, whereas the $\{C_k\}$ are the slopes of each of the individual power laws. A flat offset, $N$, is needed in order to model photon shot noise.
\subsubsection{The mode frequencies}\label{sec:mfreq}
Before we start modelling the observed ACPS we first need to select a suitable frequency interval for the purpose of our study. This interval should coincide with the frequency range in which p modes are located -- i.e.~from the frequency of the fundamental ($n$=0) mode, $\nu_{\rm f}$, up to the atmospheric acoustic cut-off frequency, $\nu_{\rm ac}$. 

We compute an estimate of the frequency of the fundamental mode by assuming that it scales with the mean stellar density and thus with the mean large frequency separation:
\begin{equation}
	\label{eq_Scaling1}
		\nu_{\rm{f}}=\nu_{{\rm{f}}\odot} \, (\Delta\nu/\Delta\nu_{\odot}) \, ,
\end{equation}
where $\nu_{{\rm{f}}\odot}\!=\!258 \: {\rm{\mu Hz}}$ and $\Delta\nu_{\odot}\!=\!135 \: {\rm{\mu Hz}}$. 

An estimate of the atmospheric acoustic cut-off frequency is calculated by assuming that it scales with the frequency of maximum amplitude \citep{KB95}:
\begin{equation}
	\label{eq_Scaling2}
		\nu_{\rm{ac}}=\nu_{{\rm{ac}}\odot} \, (\nu_{\rm{max}}/\nu_{{\rm{max}}\odot}) \, .	
\end{equation}
For the Sun one has $\nu_{{\rm{ac}}\odot}\!=\!5300 \: {\rm{\mu Hz}}$ and $\nu_{{\rm{max}}\odot}\!=\!3100 \: {\rm{\mu Hz}}$. 

We proceed with the assignment of frequency values to modes with degree $\ell\!=\!0,1,2$ (see \S\ref{sec:mheights} for an explanation of this upper limit on $\ell$) according to the asymptotic relation \citep[][]{Tassoul1980}:
	\begin{equation}
	\label{eq_Asymptotic1}
		\nu_{n\ell}=\Delta\nu \, (n+\frac{1}{2}\ell+\varepsilon)-\ell(\ell+1)D_0 \, ,	
	\end{equation}
where the quantity $\varepsilon$ is sensitive to the surface layers and was taken to be 1.5. Moreover, $D_0$ is sensitive to the sound speed gradient near the core. If Eq.~(\ref{eq_Asymptotic1}) holds exactly then it follows that\footnote{$\delta\nu_{13}$ is the spacing between adjacent modes with $\ell\!=\!1$ and $\ell\!=\!3$, and $\delta\nu_{01}$ is the amount by which $\ell\!=\!1$ modes are offset from the midpoint between the $\ell\!=\!0$ modes on either side.} $\delta\nu_{02}\!=\!6D_0$, $\delta\nu_{13}\!=\!10D_0$ and $\delta\nu_{01}\!=\!2D_0$. 

The automated \emph{Kepler} pipeline \citep{Hekker} used to supply input for the procedure described herein generates as output the smooth second order change in the large frequency separation as a function of $n$, ${\rm{d}}\Delta\nu/{\rm{d}}n$. In cases where an estimate of ${\rm{d}}\Delta\nu/{\rm{d}}n$ is available we opt for a modified version of Eq.~(\ref{eq_Asymptotic1}), which includes a second order correction:
	\begin{equation}
	\label{eq_Asymptotic2}
		\nu_{n\ell}=\Delta\nu \, (n+\frac{1}{2}\ell+\varepsilon)-\ell(\ell+1)D_0+(n-n_{\rm{max}})^2\,\frac{{\rm{d}}\Delta\nu/{\rm{d}}n}{2} \, .
	\end{equation}
The overtone with the highest power, $n_{\rm{max}}$, is given by\footnote{The {\texttt{round}}[ ] function rounds the argument to its closest integer.} ${\texttt{round}}[(\nu_{\rm{max}}-\nu_{\rm{f}})/\Delta\nu]$.	
	
For the sake of clarity we should again stress that individual modes in the observed PDS do not need to be tagged by angular degree $\ell$ prior to the fit. Assignment of frequency values to modes in the model of the PDS, according to either Eq.~(\ref{eq_Asymptotic1}) or Eq.~(\ref{eq_Asymptotic2}), is in fact done in complete ignorance of the correct mode tagging.	

\subsubsection{The mode heights}\label{sec:mheights}
The height of a multiplet component can be expressed as:
\begin{equation}
\label{eq_Ballot}
	S_{n \ell m}=\mathcal{E}_{\ell m}(i) \, S_{n \ell}=\mathcal{E}_{\ell m}(i) \, V_\ell^2 \alpha_{n \ell} \, . 
\end{equation}
The $\mathcal{E}_{\ell m}(i)$ factor represents mode visibility within a multiplet and $i$ is the inclination angle between the direction of the stellar rotation axis and the line of sight. This factor is given by \citep[][]{GizonSolanki}:     
\begin{equation}
\label{eq_ModeVisibility}
	\mathcal{E}_{\ell m}(i) = \frac{(\ell-|m|)!}{(\ell+|m|)!} \left[P_\ell^{|m|}(\cos i)\right]^2 \, ,
\end{equation}
where $P_\ell^m(x)$ are the associated Legendre functions. The quantity $V_\ell^2$ is an estimate of the geometrical visibility of the total power in a $(n,\ell)$ multiplet as a function of $\ell$, whereas $\alpha_{n\ell}$ depends mainly on the frequency and excitation mechanism. Eq.~(\ref{eq_Ballot}) is only strictly valid under one assumption: When the stellar flux is integrated over the full apparent disc, one must assume that the weighting function depends only on the distance to the disc centre. In this case, the apparent mode amplitude can effectively be separated into two factors: $\mathcal{E}_{\ell m}(i)$ and $V_\ell^2$. This assumption holds very well in the case of intensity measurements, since the weighting function is then mainly linked to the limb-darkening, whereas for velocity measurements, departures might be observed due to asymmetries in the velocity field induced by rotation (see \citealt*{BallotEtAl.2006}; \citealt{BallotEtAl.2008}, and references therein). The heights of non-radial modes are commonly defined based on the heights of radial modes according to Eq.~(\ref{eq_Ballot}), and taking into account the $V_\ell/V_0$ ratios \citep[see e.g.,][]{Bedding}. Note that $\ell\!=\!0$ modes constitute a sensible reference since they are not split by rotation. In the present case of stellar observations or when observing the Sun-as-a-star, the associated whole-disc light integration strongly suppresses high-degree modes due to the lack of spatial resolution. Stellar observations are hence mostly sensitive to high-order acoustic eigenmodes with $\ell\!\leq\!3$.

We compute the power envelope for radial modes as a function of frequency according to \citet[][]{KjeldsenEtAl.2008}. We start by subtracting a fit to the background signal from the observed PDS. The residual spectrum thus obtained is heavily smoothed over the range occupied by the p modes by convolving it with a Gaussian having a FWHM of $4\,\Delta\nu$. Finally, we multiply the smoothed, residual spectrum by $\Delta\nu/c$, where $c$ is a factor that measures the effective number of modes per order and taken to be 3.03. The height, $S_{n0}$, of a radial mode in units of power density is then given by \citep[][]{ChaplinEtAl.2008}:
\begin{equation}
\label{eq_Height}
	S_{n0}=\frac{2 A^2 \, T}{\pi T \Gamma_{n0} +2} \, ,	
\end{equation}
where $A^2$ is the total power of the mode (as determined from the power envelope for radial modes) and $T$ is the effective observational length.
\subsubsection{Setting up the model}
In order to generate the model p-mode spectrum, our modelling procedure requires the following input: The observed PDS sampled at the true resolution (i.e.~computed over a grid of uncorrelated frequency bins), a fit to the background signal, $\nu_{\rm{max}}$ and $\Delta\nu$. The gradient of the large frequency separation with $n$, ${\rm{d}}\Delta\nu/{\rm{d}}n$, is optional. This input, with the exception of the observed PDS, might be obtained from the automated \emph{Kepler} pipelines.

A common linewidth value is assigned to all the modes (a sensible assumption if the range in frequency of the spectrum being tested is not too wide). The fitted parameter $\langle\Gamma\rangle$ might then be interpreted as a mean linewidth. 

The model p-mode spectrum (including the background signal) is finally assembled according to Eq.~(\ref{eq_LimitSpectrum}), taking into account the global seismic parameters fitted in our procedure, namely, $D_0$, $\nu_{\rm{s}}$, $\langle\Gamma\rangle$ and $i$. In case we would want to allow for the effect of the windowing and statistical weighting, then we should at this stage convolve the model p-mode spectrum with the spectral window. 

\subsection{Bayesian inference}\label{sec:bayesian}

\subsubsection{Parameter estimation}
In the framework of Bayesian parameter estimation, a particular model of the observed ACPS is assumed to be true and the hypothesis space of interest relates to the values taken by the model parameters, $\bmath\Theta\!=\!\{D_0,\nu_{\rm{s}},\langle\Gamma\rangle,i\}$. These parameters are continuous, meaning that the quantity of interest is a PDF, in contrast to traditional point estimate methods. Let us state Bayes' theorem:
\begin{equation}
\label{eq_Bayes}
	p(\bmath\Theta|D,I) = \frac{p(\bmath\Theta|I) \, p(D|\bmath\Theta,I)}{p(D|I)} \, ,
\end{equation}
where $D$ represents the available data and the prior information is represented by $I$. $p(\bmath\Theta|I)$ is called the \emph{prior probability}, whereas $p(\bmath\Theta|D,I)$ is called the \emph{posterior probability}. The \emph{likelihood function} is represented by $p(D|\bmath\Theta,I)$. $p(D|I)$, the \emph{global likelihood}, simply plays the role of a normalisation constant. 

The procedure of \emph{marginalisation} makes it possible to derive the marginal posterior PDF for a subset of parameters $\bmath\Theta_{\rm{A}}$, by integrating out the remaining parameters $\bmath\Theta_{\rm{B}}$, the so-called \emph{nuisance parameters}:
\begin{equation}
\label{eq_Marginalization}
	p(\bmath\Theta_{\rm{A}}|D,I) = \int p(\bmath\Theta_{\rm{A}},\bmath\Theta_{\rm{B}}|D,I) \, \mathrm{d}\bmath\Theta_{\rm{B}} \, .
\end{equation}
Furthermore, assuming that the prior on $\bmath\Theta_{\rm{A}}$ is independent of the prior on the remaining parameters, then by applying the product rule we have:
\begin{equation}
\label{eq_Priors}
	p(\bmath\Theta_{\rm{A}},\bmath\Theta_{\rm{B}}|I)\!=\!p(\bmath\Theta_{\rm{A}}|I)p(\bmath\Theta_{\rm{B}}|\bmath\Theta_{\rm{A}},I)\!=\!p(\bmath\Theta_{\rm{A}}|I)p(\bmath\Theta_{\rm{B}}|I) \, .
\end{equation} 

The main advantage of the Bayesian framework when compared to a frequentist approach is the ability to incorporate relevant prior information through Bayes' theorem and to evaluate its effect on our analysis \citep[see e.g.,][for possible applications of Bayesian probability theory in Astrophysics]{Brewer}. 

\subsubsection{The likelihood function}
We would like to specify the likelihood function, i.e.~the joint PDF for the data sample. We know that, for a given frequency lag $L_i \in [L_{\rm{min}},L_{\rm{max}}]$:
\begin{equation}
\label{eq_Likelihood1}
	{\rm{ACPS_{o}}}(L_i)={\rm{ACPS_{m}}}(L_i)+e_i \, ,
\end{equation}
where ${\rm{ACPS_{o}}}$ and ${\rm{ACPS_{m}}}$ are the observed and model ACPS, respectively. The error term $e_i$ follows a Gaussian\footnote{Each point in the ACPS is the sum of a large number of independent and identically distributed random variables. Therefore, the Central Limit Theorem applies and the distribution of $e_i$ should approach the normal distribution.} distribution with zero mean and standard deviation $\sigma_i$. Hence, assuming that the model is deterministic, i.e.~true, we can write
\begin{eqnarray}
\label{eq_Likelihood2}
	f\left[{\rm{ACPS_{o}}}(L_i)\right]=\frac{1}{\sigma\sqrt{2\pi}} \exp\left\{-\frac{e_i^2}{2\sigma^2}\right\}= \nonumber \\
	\frac{1}{\sigma\sqrt{2\pi}} \exp\left\{-\frac{\left[{\rm{ACPS_{o}}}(L_i)-{\rm{ACPS_{m}}}(L_i)\right]^2}{2\sigma^2}\right\} \, ,
\end{eqnarray}
where $f\left[{\rm{ACPS_{o}}}(L_i)\right]$ is the probability density that the observed ACPS takes a particular value ${\rm{ACPS_{o}}}(L_i)$ at a frequency lag $L_i$. Notice that we are assuming $\sigma_i$ to be constant over the whole range spanned by $L_i$. 

By ignoring the effect of correlation between points in the ACPS we arrive at the following expression for the likelihood function:
\begin{equation}
\label{eq_Likelihood3}
	\mathcal{L}(\bmath\Theta)\equiv p(D|\bmath\Theta,I)=\prod_i f\left[{\rm{ACPS_{o}}}(L_i)\right] \, .
\end{equation}
While the effect of correlation clearly must be present, the fact that we ignore it can be shown to only affect attempted error calculations and not the fitted values themselves \citep[see][and references therein]{FletcherEtAl.2006}. This does not constitute a problem in the present case, since we will be employing a MCMC sampler in order to obtain the marginal posteriors for each of the model parameters, from which we can estimate the uncertainties.

\subsection{Markov chain Monte Carlo (MCMC)}\label{sec:MCMC}  
After inspection of Eq.~(\ref{eq_Marginalization}), the need for a mathematical tool that is able to efficiently evaluate the multi-dimensional integrals required in the computation of the marginal posteriors becomes clear. This constitutes the principal rationale behind the method known as \emph{Markov chain Monte Carlo}, which aims at drawing samples from the \emph{target distribution}, $p(\bmath\Theta|D,I)$, by constructing a pseudo-random walk in model parameter space. Such a pseudo-random walk is achieved by generating a Markov chain, whereby a new sample, $\bmath\Theta_{t+1}$, depends solely on the previous sample, $\bmath\Theta_{t}$, in accordance with a time-independent quantity called the \emph{transition kernel}, $p(\bmath\Theta_{t+1}|\bmath\Theta_{t})$:
\begin{eqnarray}
\label{eq_Kernel}
	p(\bmath\Theta_{t+1}|\bmath\Theta_{t})= \nonumber \\
	q(\bmath\Theta_{t+1}|\bmath\Theta_{t}) \min\left[1,\frac{p(\bmath\Theta_{t+1}|D,I)}{p(\bmath\Theta_{t}|D,I)} \frac{q(\bmath\Theta_{t}|\bmath\Theta_{t+1})}{q(\bmath\Theta_{t+1}|\bmath\Theta_{t})}\right] \, ,
\end{eqnarray}
where $q(\bmath\Theta_{t+1}|\bmath\Theta_{t})$ is a \emph{proposal distribution} centred on $\bmath\Theta_{t}$. After a \emph{burn-in} phase, $p(\bmath\Theta_{t+1}|\bmath\Theta_{t})$ is able to generate samples of $\bmath\Theta$ with a probability density converging on the target distribution. The algorithm that we employ in order to generate a Markov chain was initially proposed by \citet[][]{Metropolis}, and subsequently generalised by \citet[][]{Hastings}, this latter version being commonly referred to as the \emph{Metropolis--Hastings algorithm}.

The Metropolis--Hastings MCMC algorithm just outlined above runs the risk of becoming stuck in a local maximum of the target distribution, thus failing to fully explore all regions in parameter space containing significant probability. A way of overcoming this difficulty is to employ \emph{parallel tempering} \citep[see e.g.,][]{EarlDeem}, whereby a discrete set of progressively flatter versions of the target distribution is created by introducing a \emph{temperature parameter}, $\mathcal{T}$. In practice, use is made of its reciprocal, $\beta\!=\!1/\mathcal{T}$, referred to as the \emph{tempering parameter}. By modifying Eq.~(\ref{eq_Bayes}), we generate the tempered distributions as follows:
\begin{equation}
\label{eq_ParallelTempering}
	p(\bmath\Theta|D,\beta,I) = C \, p(\bmath\Theta|I) \, p(D|\bmath\Theta,I)^\beta \, , \quad 0 < \beta \leq 1 \, ,
\end{equation}
where $C$ is a constant. For $\beta\!=\!1$, we retrieve the target distribution, also called the \emph{cold sampler}, whereas for $\beta\!<\!1$, the hotter distributions are effectively flatter versions of the target distribution. By running such a set of cooperative chains in parallel and by further allowing for the swap of their respective parameter states, we enable the algorithm to sample the target distribution in a way that allows for both the investigation of its overall features (low-$\beta$ chains) and the examination of the fine details of a local maximum (high-$\beta$ chains). See Appendix \ref{appendix} for the current version of the parallel tempering Metropolis--Hastings algorithm written in pseudocode.

Moreover, based on a statistical \emph{control system} (CS) similar to the one described in \citet[][]{GregoryBook}, we automate the process of calibration of the Gaussian proposal $\sigma_{\rm{CS}}$-values, which specify the direction and step size in parameter space when proposing a new sample to be drawn. The optimal choice of $\{\sigma_{\rm{CS}}\}$ is closely related to the average rate at which proposed state changes are accepted, the so-called \emph{acceptance rate}. The control system makes use of an error signal to steer the selection of the $\sigma_{\rm{CS}}$-values during the burn-in stage of a single parallel tempering MCMC run, acting independently on each of the tempered chains. The error signal is proportional to the difference between the current acceptance rate and the target acceptance rate. As soon as the error signal for each of the tempered chains is less than a measure of the statistical fluctuation expected for a zero mean error, the control system is turned off and the algorithm switches to the standard parallel tempering MCMC. 

\section{Application to Simulated Data: The Solar Twin Boris}\label{sec:results}
We display the results obtained when applying the ACPS-modelling procedure to the particular case of Boris, an artificial main-sequence star created in the framework of the AsteroFLAG group for the purpose of conducting hare and hounds. Briefly, Boris is what we might call a solar analog ($L=1.00\,{\rm{L}}_{\odot};\,T_{\rm{eff}}=5778\:{\rm{K}};\,R=1.00\,{\rm{R}}_{\odot}$). We refer the reader to \citet[][]{AsteroFLAG1} and \citet[][]{AsteroFLAG2} for further insight into AsteroFLAG's activities.

\subsection{Experimental setup}
Boris was assumed to have an apparent visual magnitude of either $V\!=\!8$ or $V\!=\!9$ or $V\!=\!10$ (these values correspond to the bright end of the nominal apparent magnitude target range for \emph{Kepler}). In each case, 250 realisations of a PDS were generated. The goal was to assess both the accuracy and the precision of the several Bayesian summary statistics of the fitted parameters, namely, the MAP (Maximum \emph{A Posteriori}), the median, the mean and the marginal posterior mode. These power density spectra were directly generated in Fourier space and include contributions arising from p modes, granulation, activity, photon shot noise and instrumental noise. They correspond to 30-day long time series with a 60-second cadence, as it would be expected for \emph{Kepler}'s high-cadence survey targets. The stellar inclination angle, $i$, entering the simulations was allowed to vary and was drawn from a probability distribution that assumes random orientations -- $\sin(i)$. Reference values are given for the other model parameters by: $D_0^{\rm{ref}}\!=\!1.43 \: {\rm{\mu Hz}}$, $\nu_{\rm{s}}^{\rm{ref}}\!=\!\nu_{{\rm{s}}\odot}\!=\!0.4 \: {\rm{\mu Hz}}$ and $\langle\Gamma\rangle^{\rm{ref}}\!=\!1.95 \: {\rm{\mu Hz}}$. All the remaining relevant quantities that enter the simulations were assigned `solar' values. 

A prototype of the Birmingham--Sheffield Hallam automated pipeline \citep[][]{Hekker} was then run on the simulated power density spectra in order to retrieve useful input for the ACPS-modelling procedure, to be specific, a fit to the background signal, $\Delta\nu$ and ${\rm{d}}\Delta\nu/{\rm{d}}n$. The frequency interval of interest (for purposes of the computation of the ACPS) ranges from 2100 to $4500 \: {\rm{\mu Hz}}$ and $\nu_{\rm{max}}$ has been fixed at $3300 \: {\rm{\mu Hz}}$.

\begin{figure}
\centering
\begin{tabular}{c c}
\epsfig{file=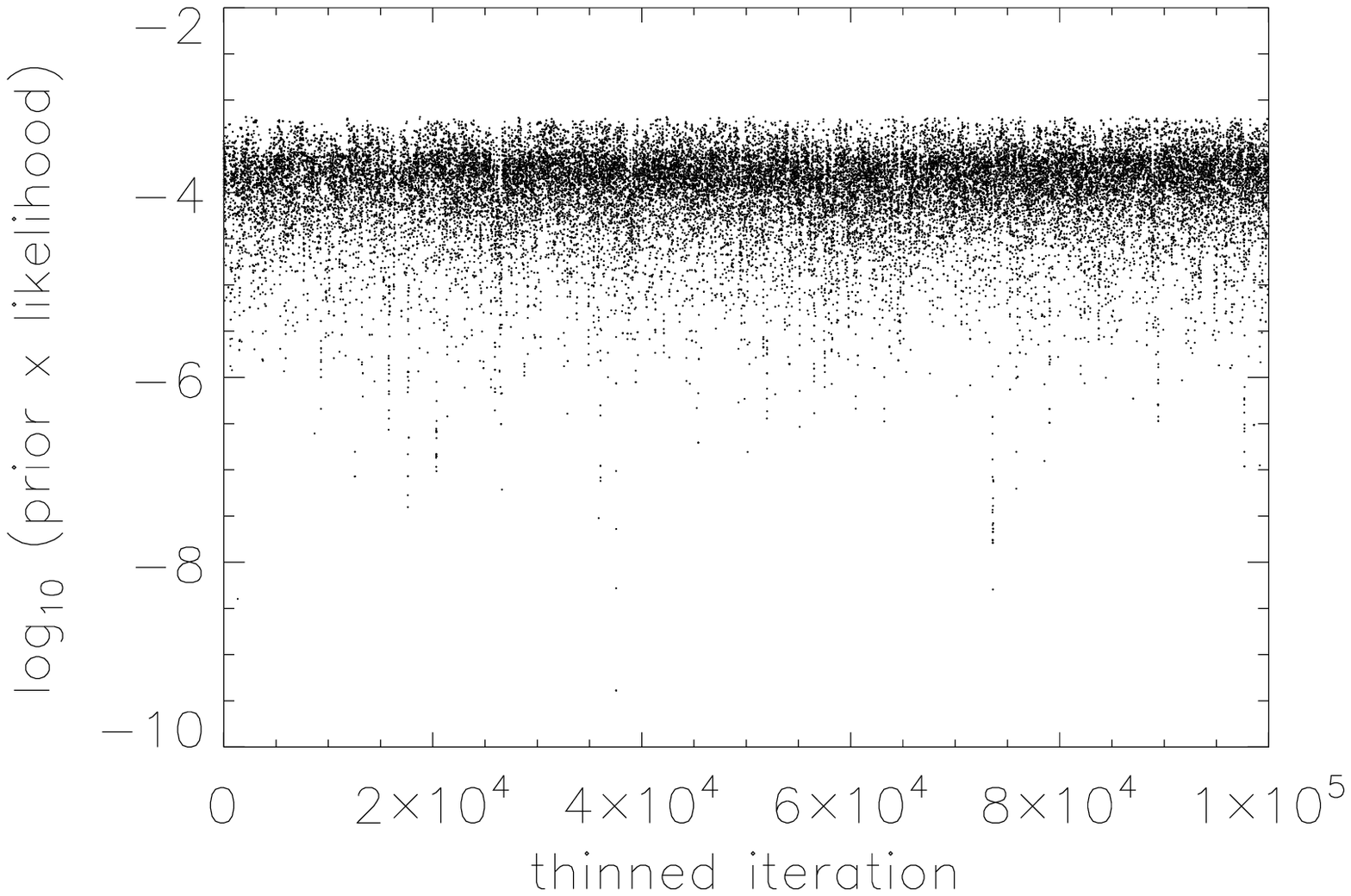,width=0.48\linewidth,clip=} & \epsfig{file=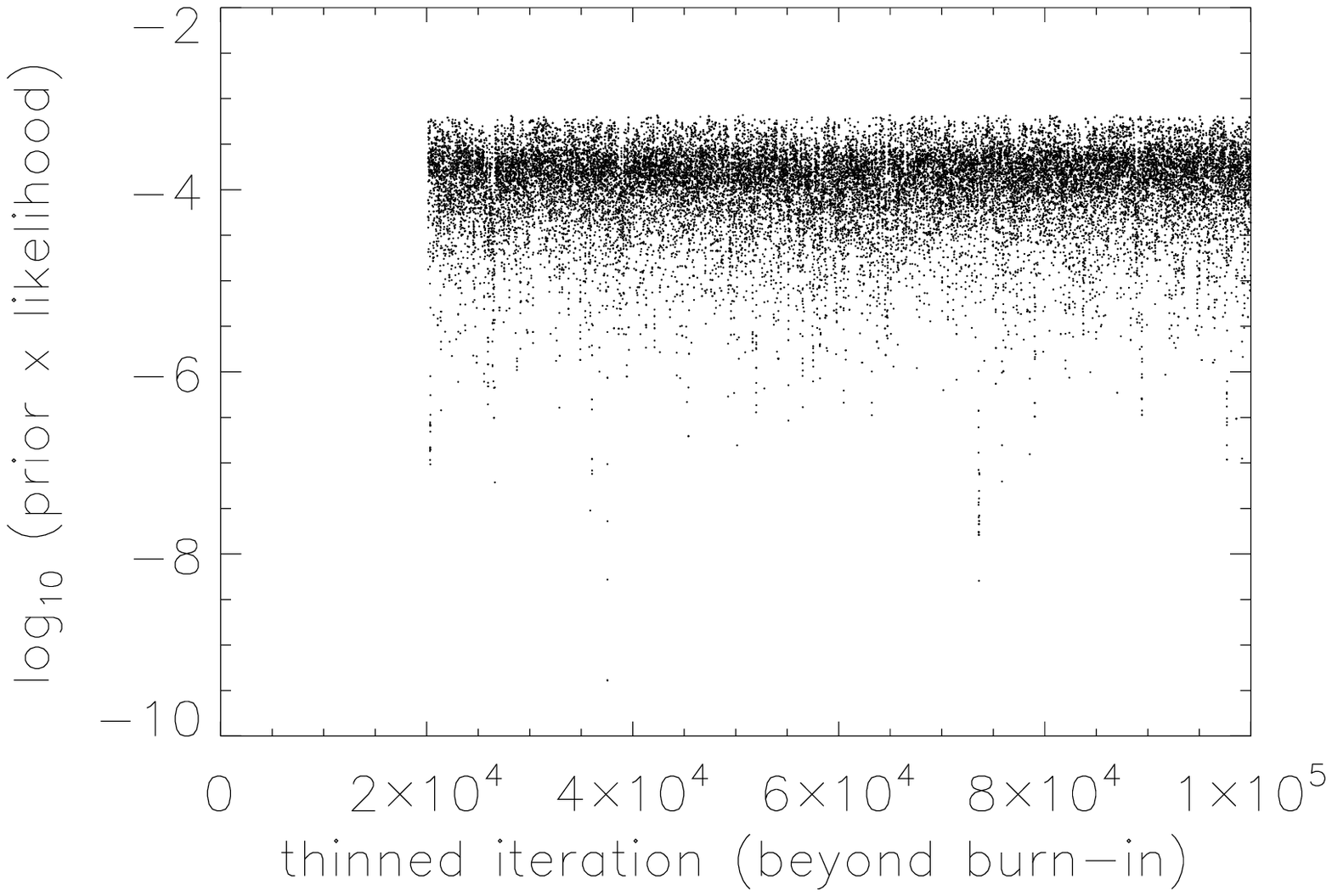,width=0.48\linewidth,clip=} \\
\epsfig{file=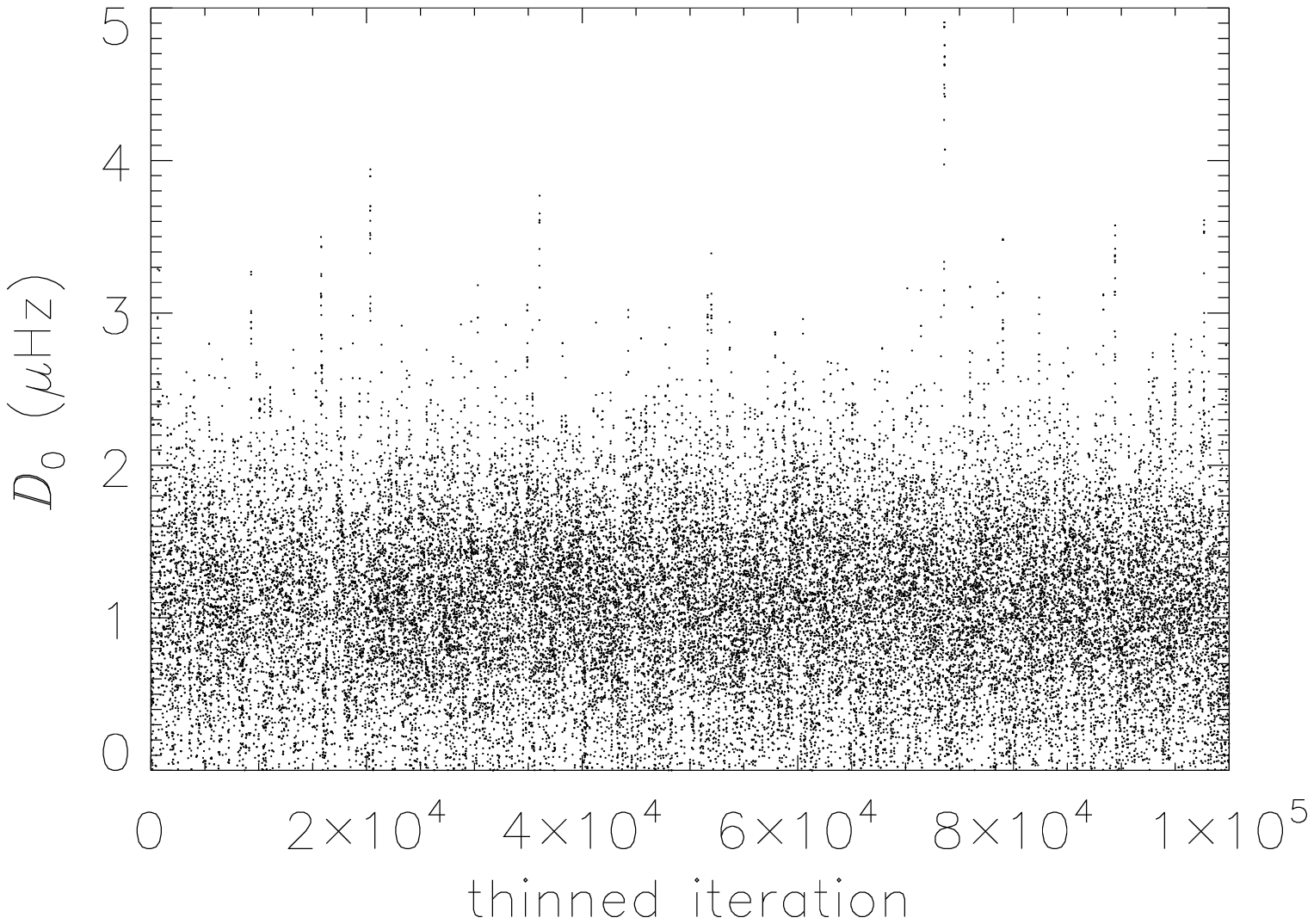,width=0.48\linewidth,clip=} & \epsfig{file=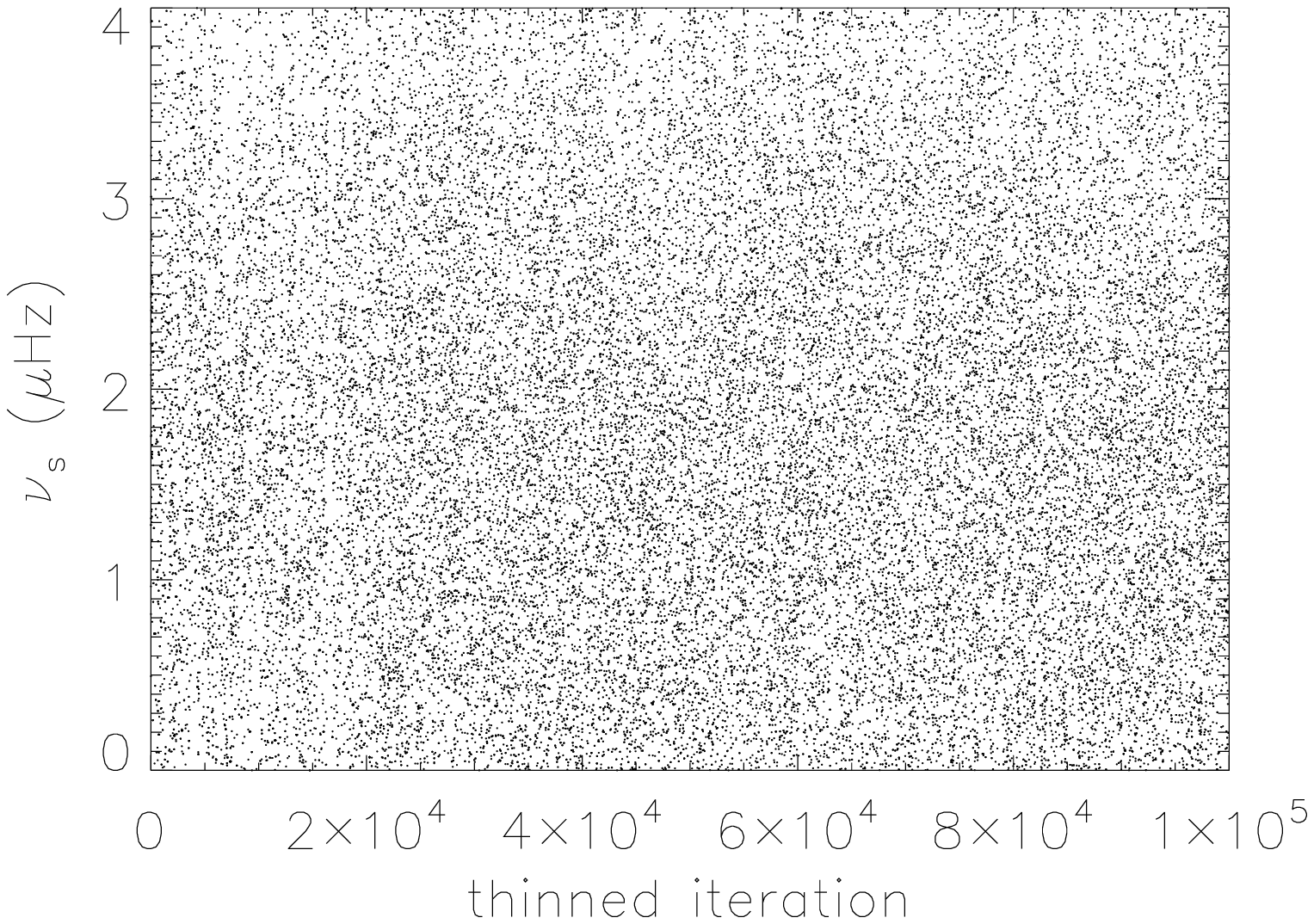,width=0.48\linewidth,clip=} \\
\epsfig{file=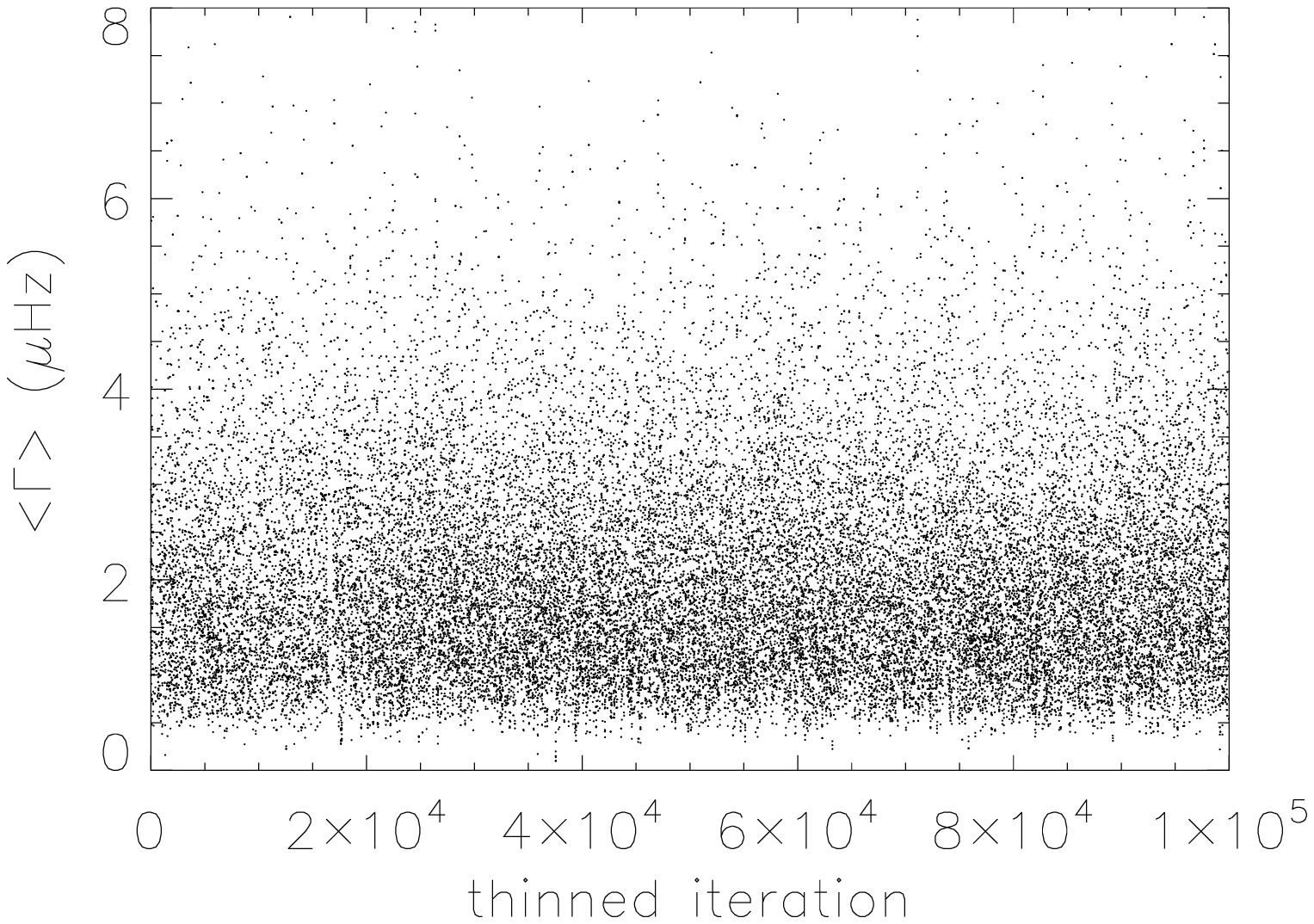,width=0.48\linewidth,clip=} & \epsfig{file=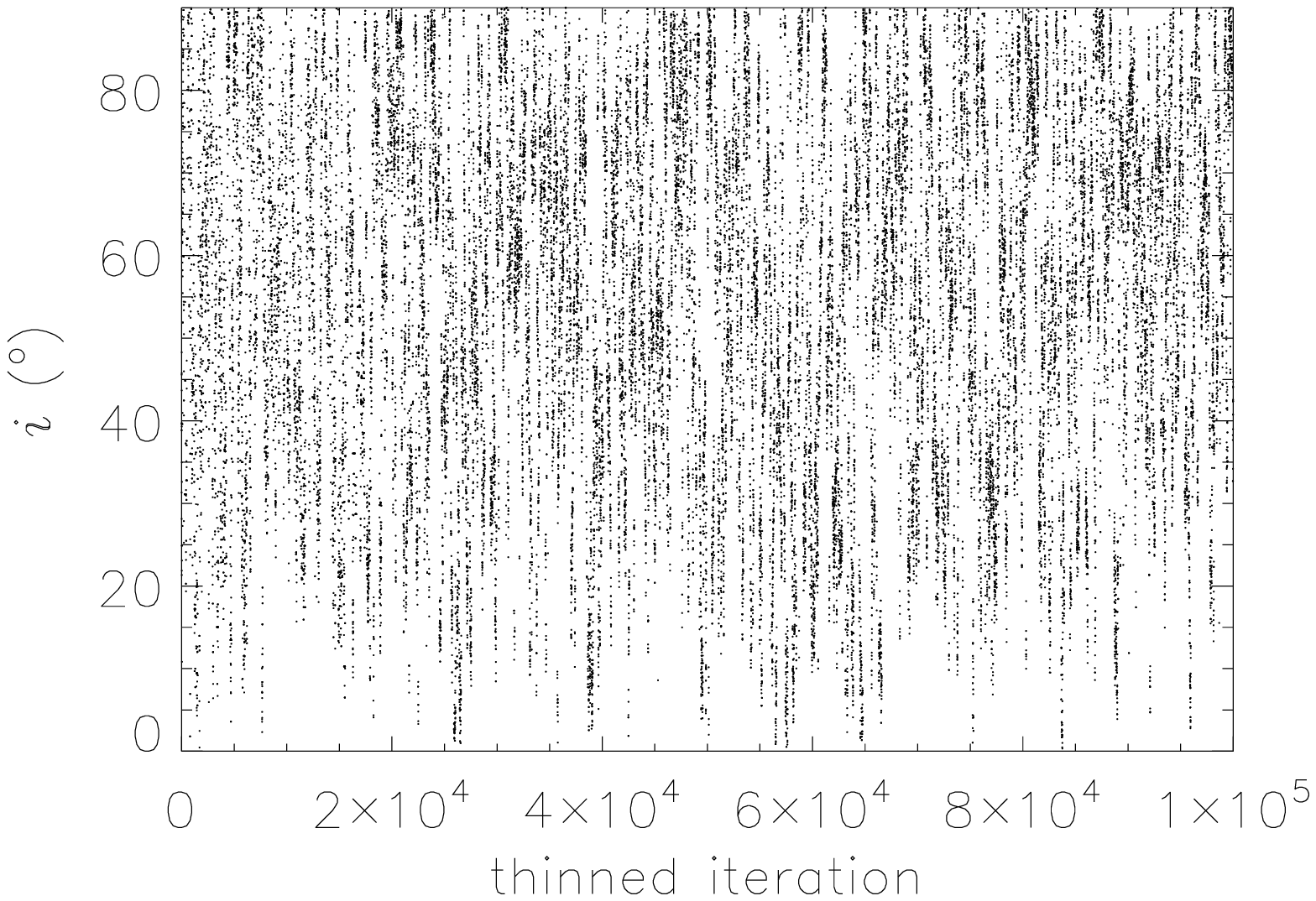,width=0.48\linewidth,clip=} 
\end{tabular}
\caption{Sequence of $1\times10^5$ samples (at a thinning interval of 1) drawn from the target distribution ($\beta\!=\!1$) by a Metropolis--Hastings Markov chain Monte Carlo. \emph{Top Panels:} Behaviour of $\log_{10} \left[p(\bmath\Theta|I)\,\mathcal{L}(\bmath\Theta)\right]$ as a function of the iteration number (a means of visually inspecting the convergence of the MCMC). \emph{Middle and Bottom Panels:} Behaviour of model parameter values as a function of the iteration number.}
\label{fig_Diagnose1}
\end{figure}

\begin{figure}
     \resizebox{\hsize}{!}{\includegraphics{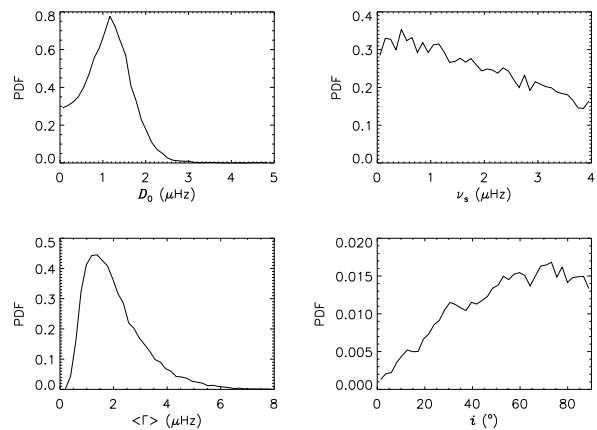}} 
     \caption{Binned marginal posterior PDFs of the model parameters. These are built after assembling the corresponding samples (beyond burn-in) depicted in the middle and bottom panels of Fig.~\ref{fig_Diagnose1}. Based on these marginal posteriors we may then evaluate the several Bayesian summary statistics for each of the model parameters.}
     \label{fig_Diagnose3}
 \end{figure}

Figs.~\ref{fig_Diagnose4}, \ref{fig_Diagnose1} and \ref{fig_Diagnose3} display the typical graphical output of the ACPS-modelling procedure resulting from the analysis of a single realisation of the PDS of Boris (with $V\!=\!8$ and a random orientation characterised by $i\!=\!32.4\,^{\circ}$). Note that throughout the analysis of the full set of realisations of the power spectrum, uniform priors have been imposed on $D_0$, $\nu_{\rm{s}}$ and $\langle\Gamma\rangle$, whereas $p(i|I)\!=\!\sin(i)$.

\subsection{On retrieving $D_0$ and $\langle\Gamma\rangle$}
\begin{figure}
\centering
\begin{tabular}{c}
\epsfig{file=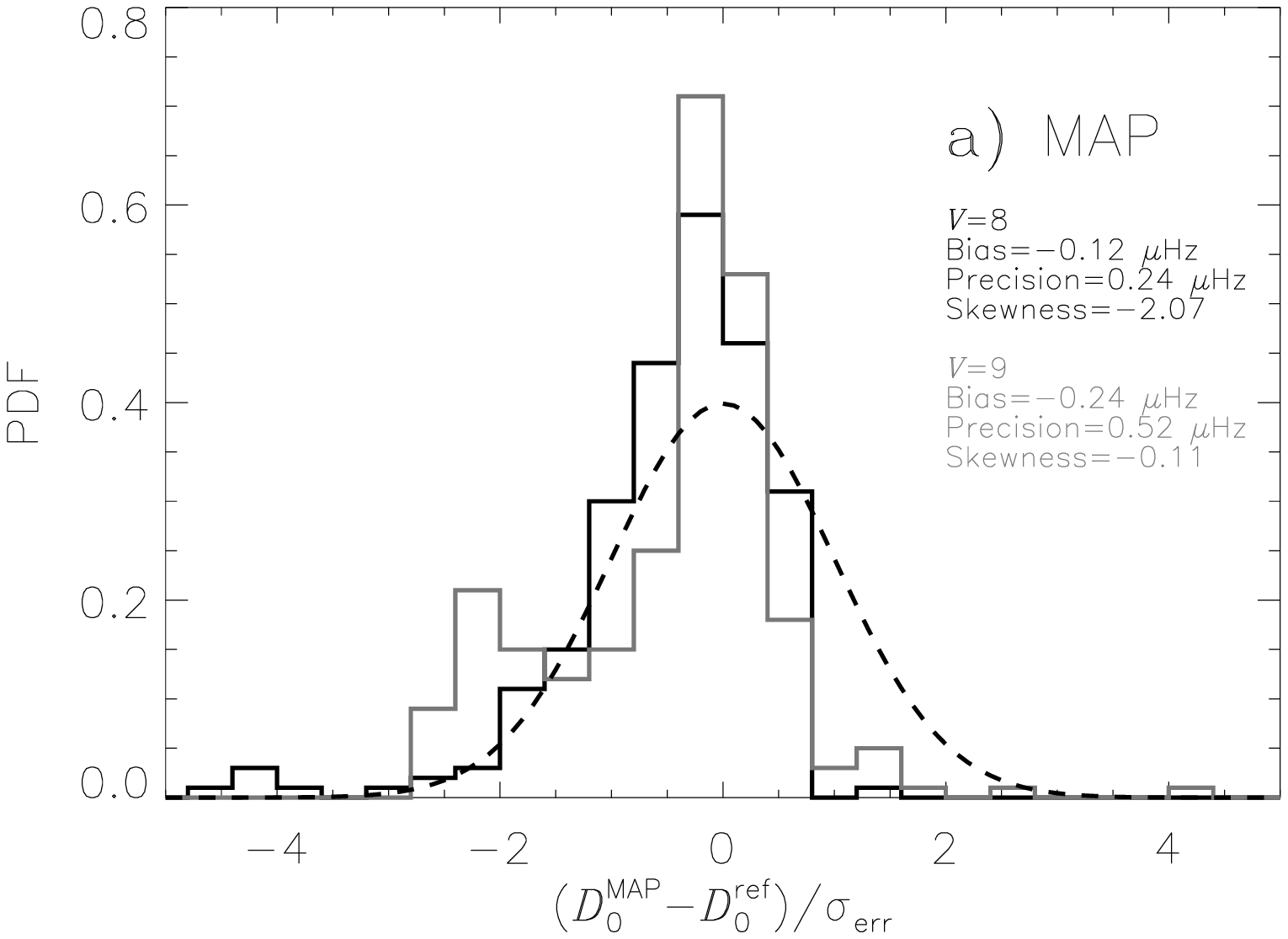,width=0.85\linewidth,clip=} \\
\epsfig{file=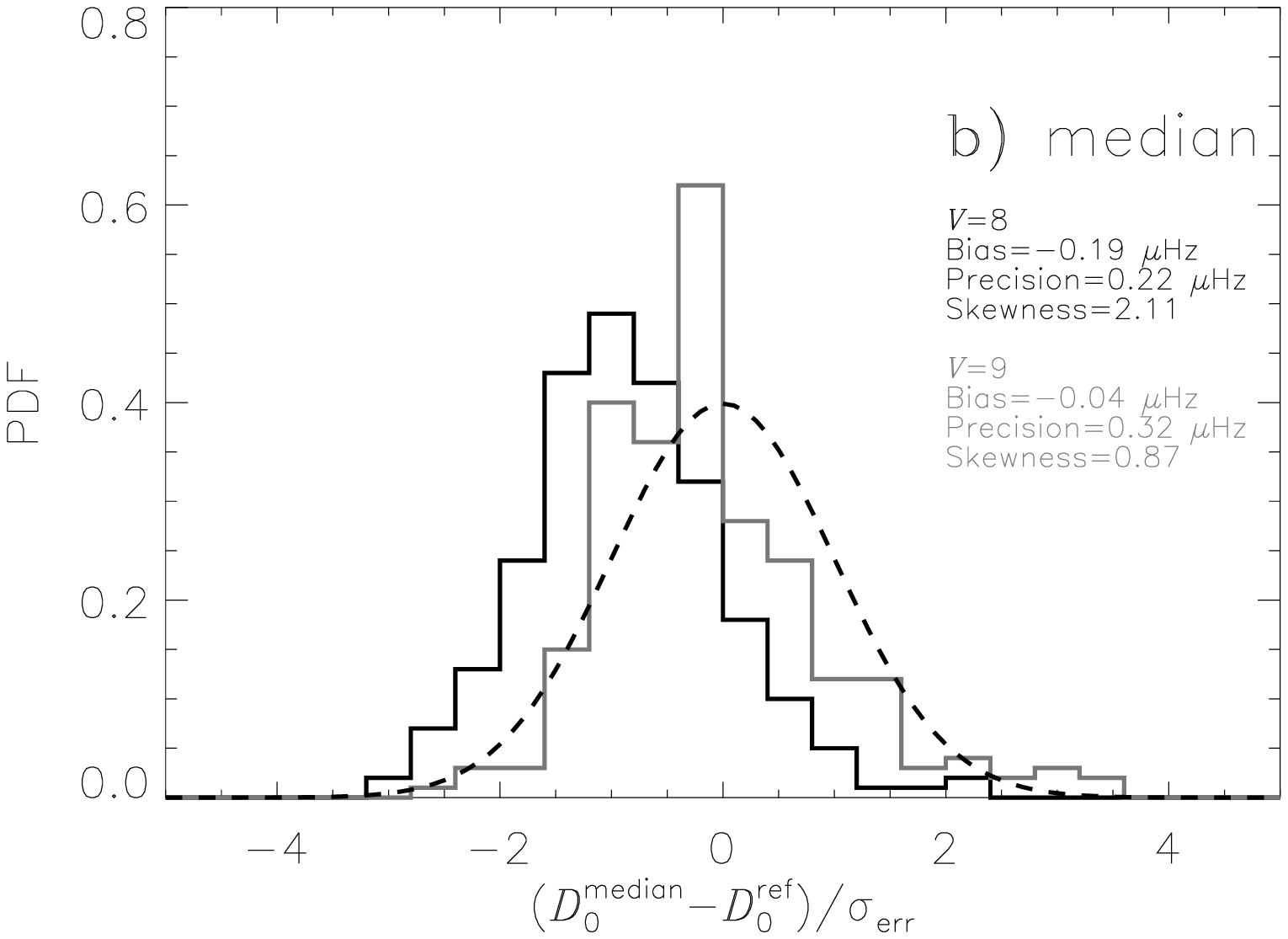,width=0.85\linewidth,clip=} \\
\epsfig{file=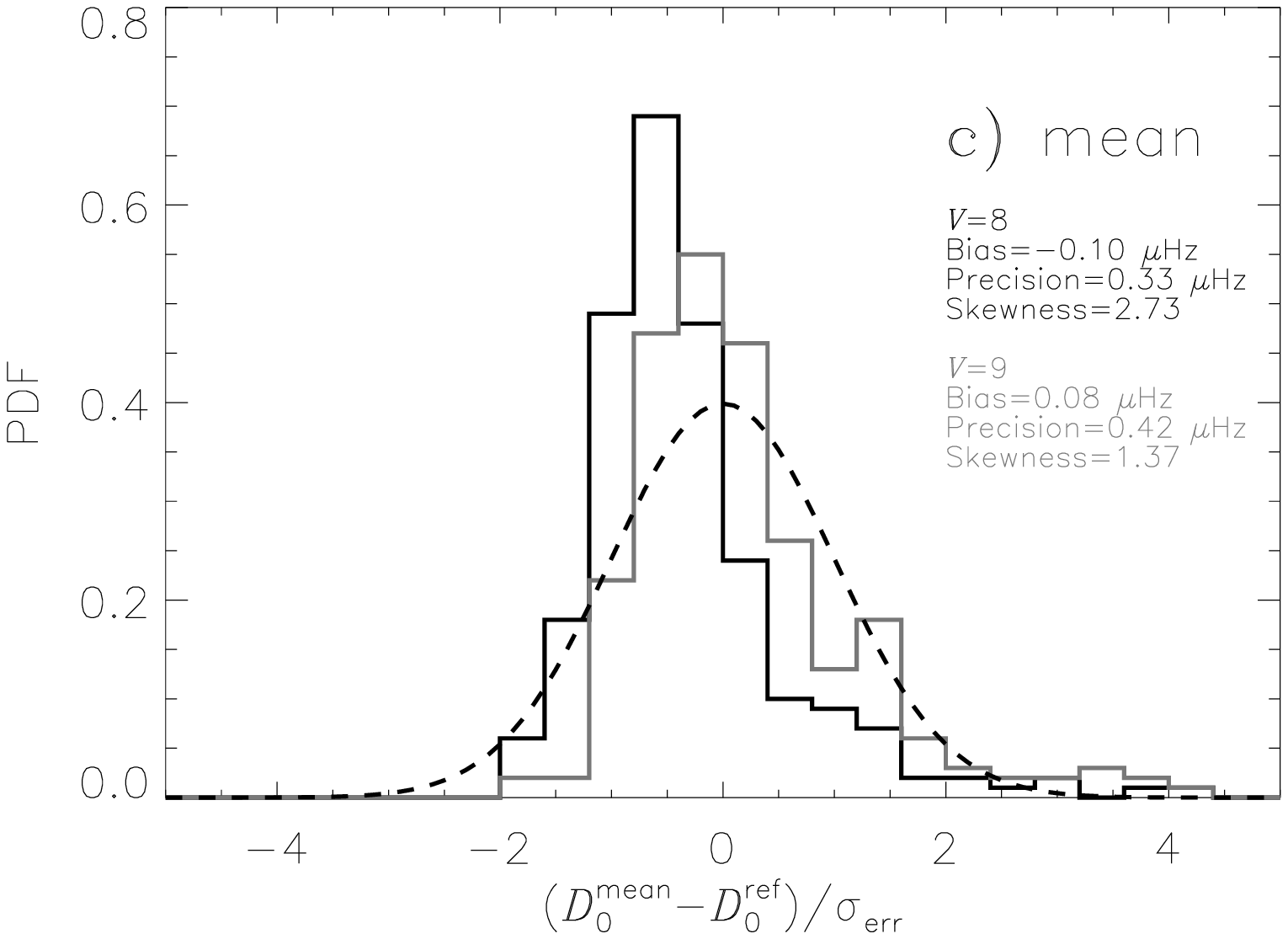,width=0.85\linewidth,clip=} \\
\epsfig{file=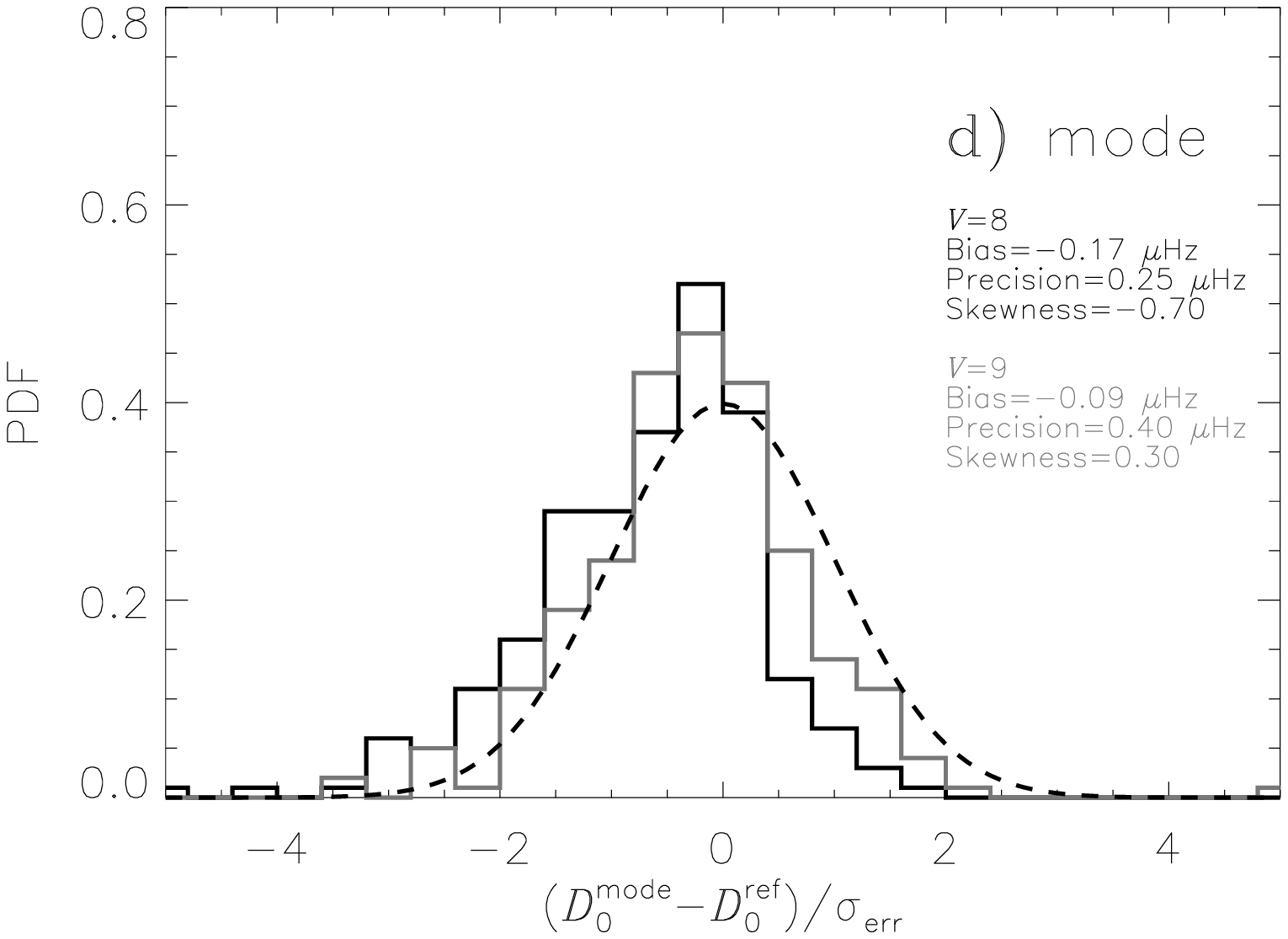,width=0.85\linewidth,clip=} 
\end{tabular}
\caption{PDFs of the actual errors in the determination of $D_0$, which in turn have been normalised after division by their standard deviation, $\sigma_{\rm{err}}$. Several Bayesian summary statistics have been considered: a) the MAP, b) the median, c) the mean and d) the mode. The solid black line corresponds to the set with $V\!=\!8$, whilst the solid grey line corresponds to the set with $V\!=\!9$. The dashed line represents the standard normal distribution.}
\label{fig_D0}
\end{figure}

\begin{figure}
\centering
\begin{tabular}{c}
\epsfig{file=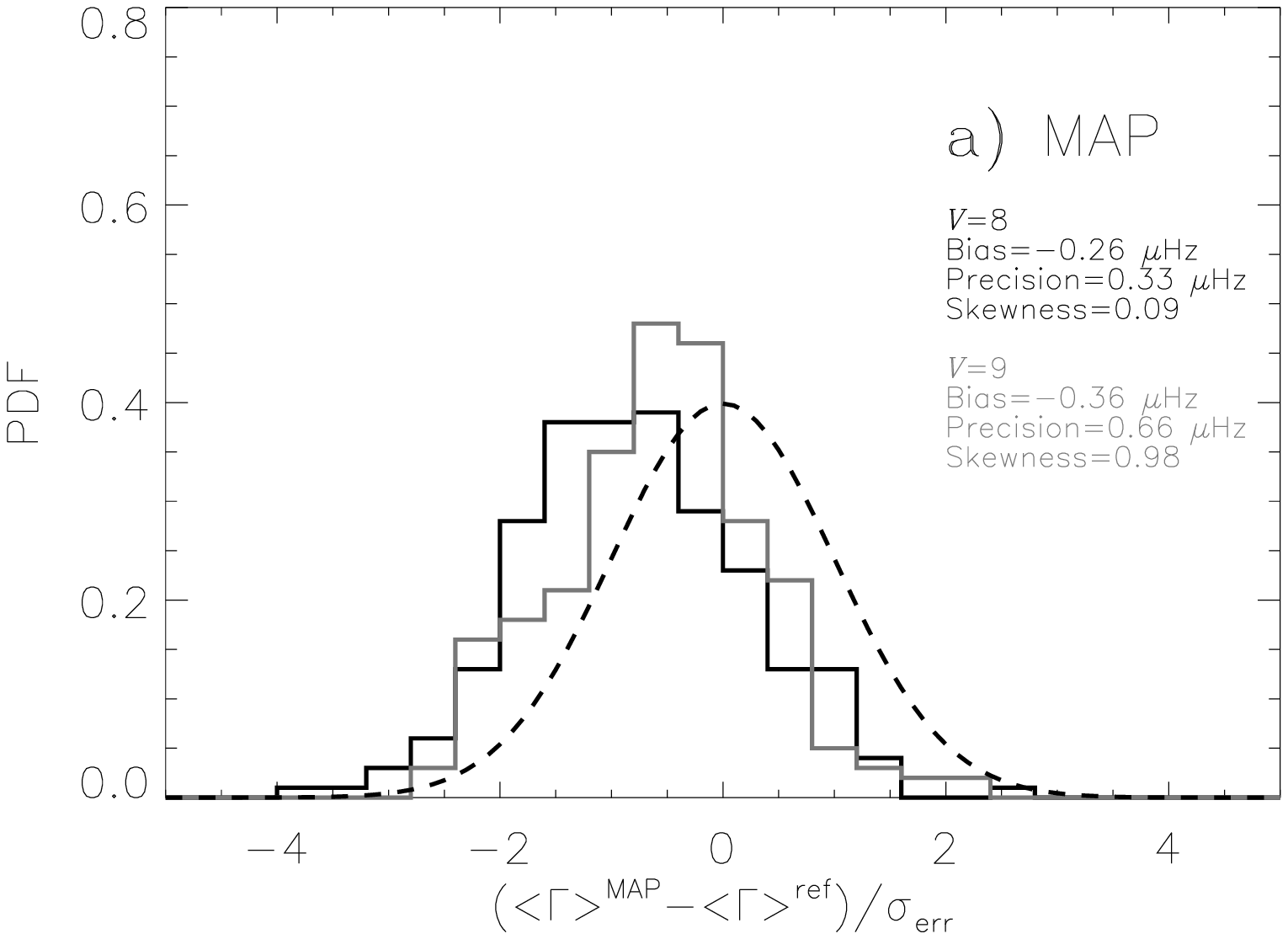,width=0.85\linewidth,clip=} \\ 
\epsfig{file=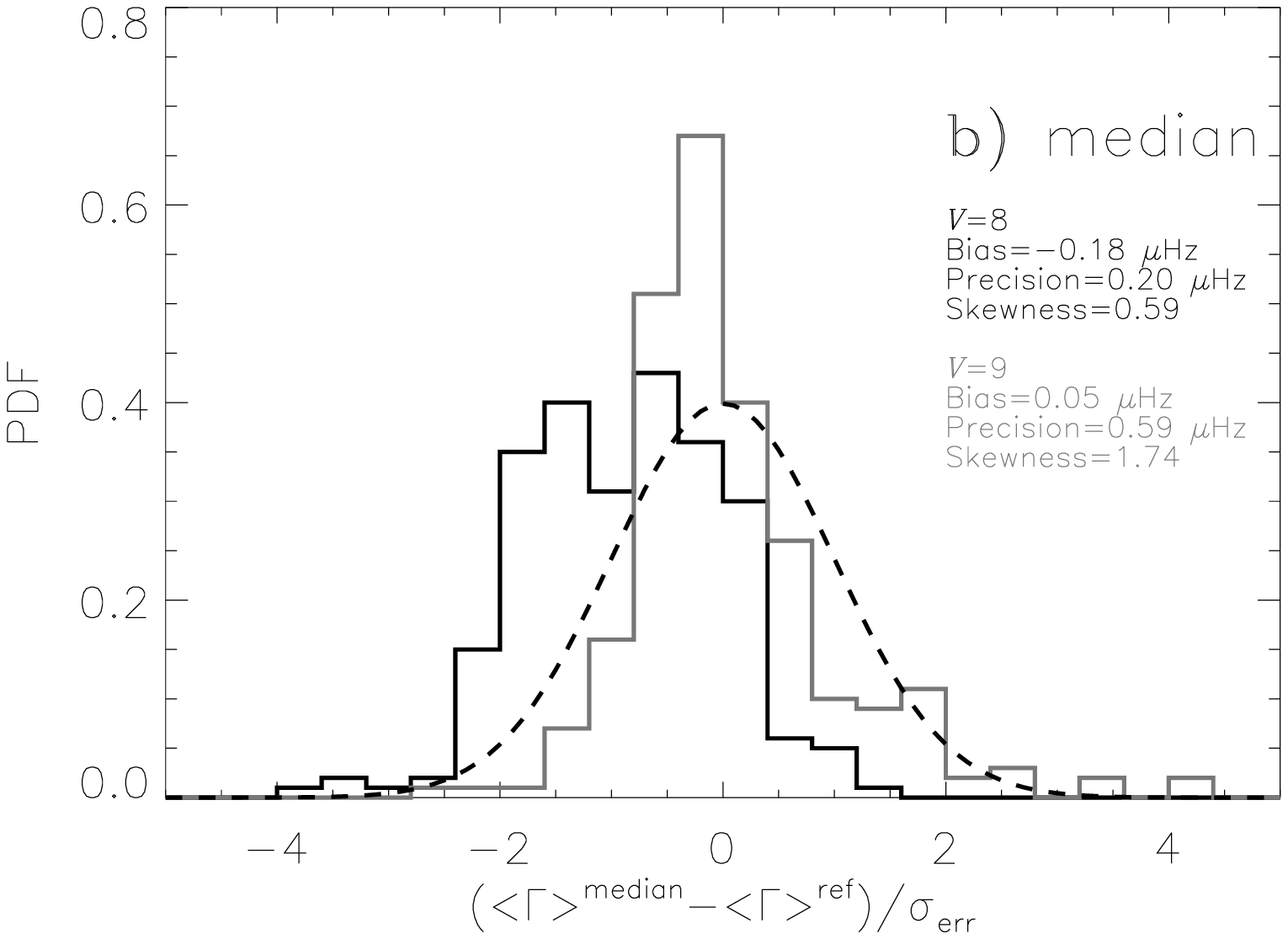,width=0.85\linewidth,clip=} \\
\epsfig{file=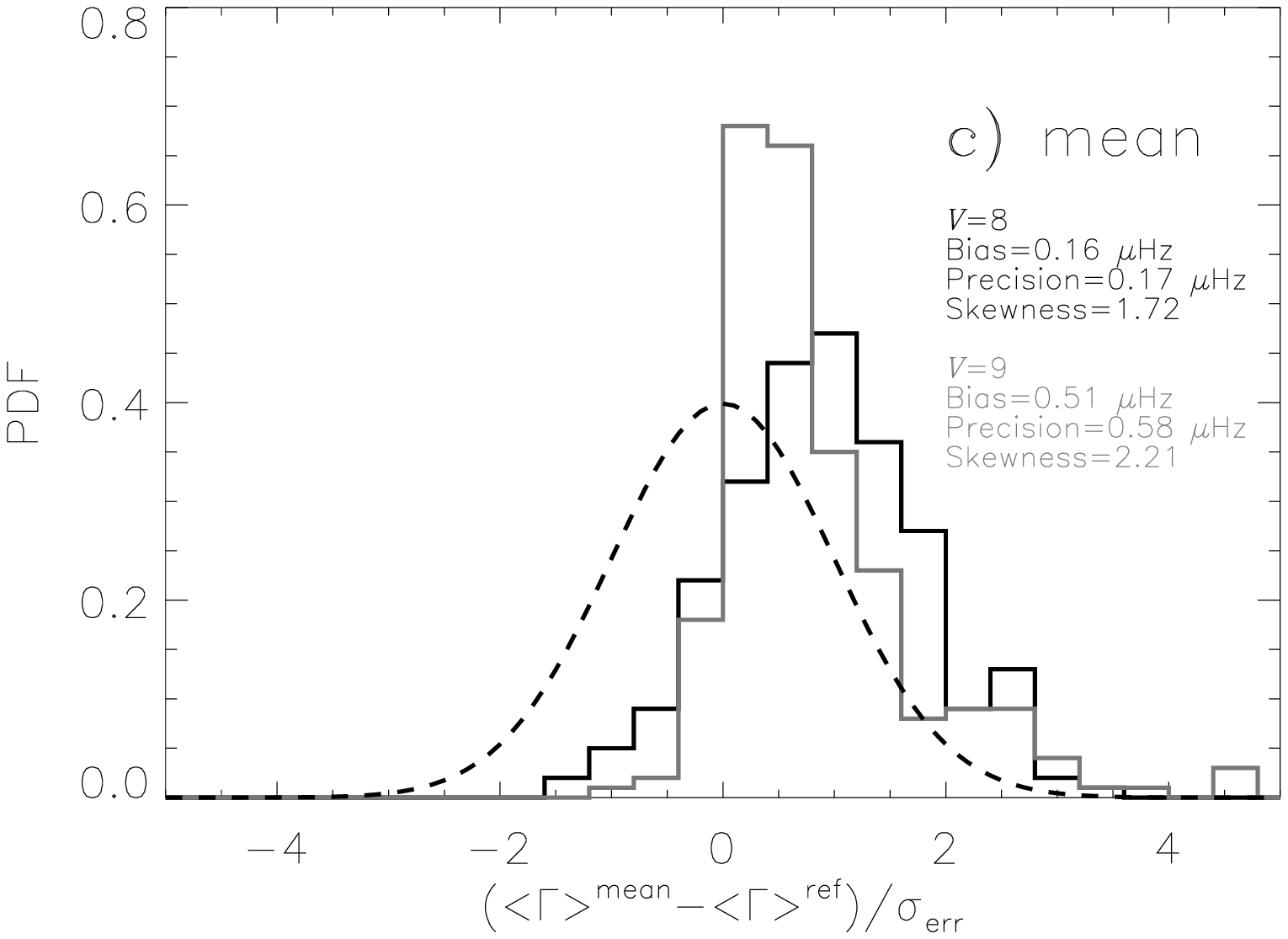,width=0.85\linewidth,clip=} \\
\epsfig{file=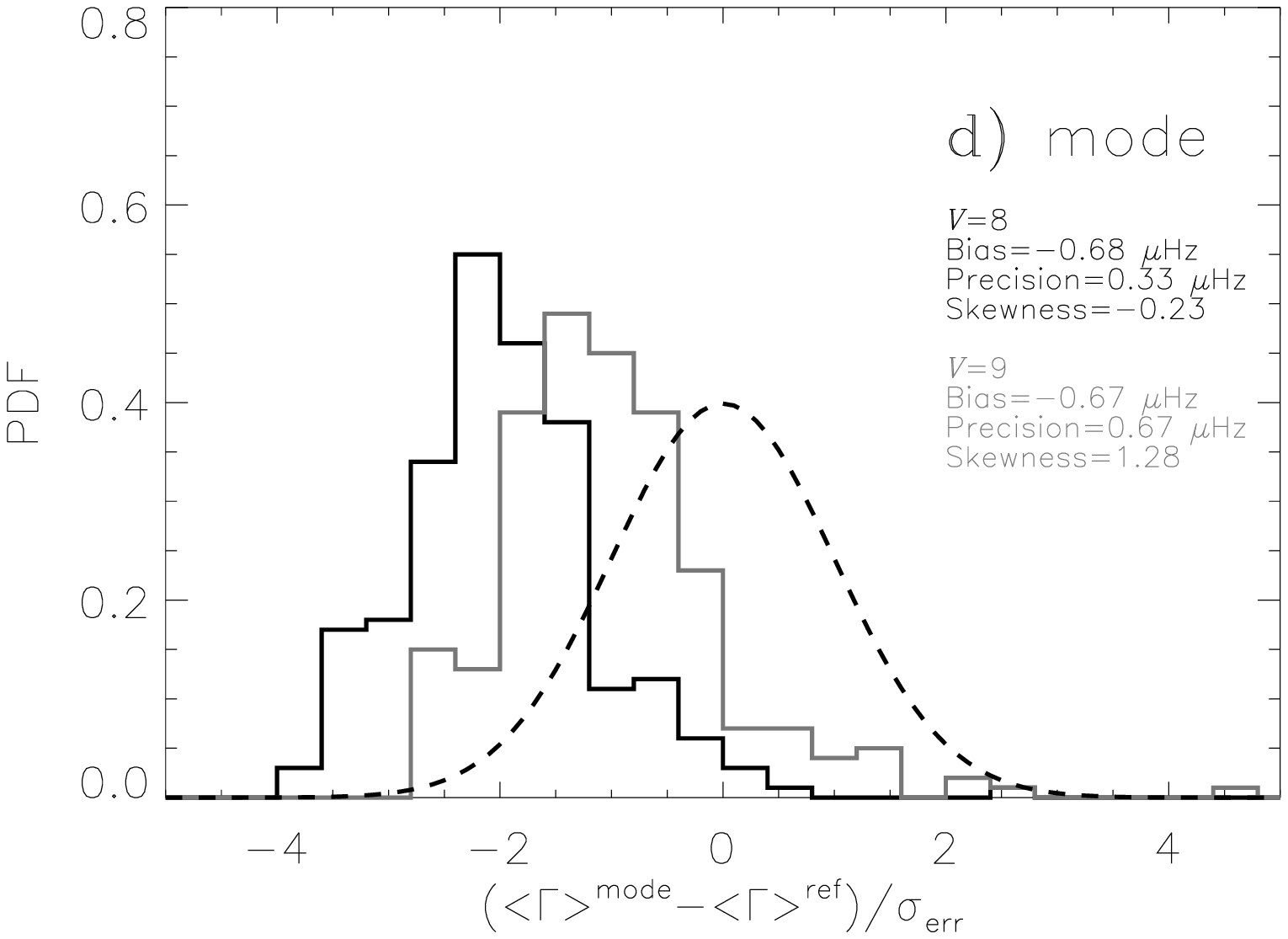,width=0.85\linewidth,clip=}
\end{tabular}
\caption{The same as in Fig.~\ref{fig_D0} but now regarding the estimation of $\langle\Gamma\rangle$.}
\label{fig_Gamma}
\end{figure}

Figs.~\ref{fig_D0} and \ref{fig_Gamma} result from the application of the ACPS-modelling procedure to the full set of realisations of the power spectrum (250 realisations for each value of $V$), allowing one to assess the potential biases of the Bayesian summary statistics of the fitted parameters $D_0$ and $\langle\Gamma\rangle$, respectively. The precision associated with these same estimators is also indicated in the plots. 

When choosing a summary statistic to be used, we should make sure that it is representative of the marginal probability distribution of the parameter in question, a point that all the summary statistics considered seem to satisfy. The marginal posterior PDFs of $D_0$ and $\langle\Gamma\rangle$ are asymmetric (see Fig.~\ref{fig_Diagnose3}) and well modelled by \citep[][]{Fraillon}:
\begin{equation}
\label{eq_Fraillon}
	f(x)=K(x/s)^{\alpha} \exp\left[(x/s)^{\alpha+1}\right] \, ,
\end{equation}
where $K$ is a normalisation constant, $s$ is an adjustable parameter describing the shape of the distribution and $\alpha$ defines the type of distribution ($\alpha\!=\!0$ for a Boltzmann law and high values of $\alpha$ for a Gaussian law). Furthermore, the set of summary parameter values should provide a good fit to the data, the quality of which might be assessed by computing the rms of the residuals. This latter criterion led us to adopt the MAP as our choice of summary statistic. 

It turns out from Figs.~\ref{fig_D0} and \ref{fig_Gamma} that the different estimators, although equally robust, present biases whose magnitude and sign vary with the estimator being considered, a direct consequence of the asymmetry of the marginal posterior PDFs of $D_0$ and $\langle\Gamma\rangle$. A way of overcoming these inherent biases would be to present these values normalised to the homologous values obtained after performing a similar analysis on Sun-as-a-star data covering an equivalent range in frequency (i.e.~scaling by the ratio of the acoustic cut-off frequencies of the Sun and the target in question).

The MAP summary statistic allows one to determine $D_0$ with a relative error of 8 per cent and 17 per cent for $V\!=\!8$ and $V\!=\!9$, respectively. On the other hand, determination of $\langle\Gamma\rangle$ is accomplished with a relative error of 13 per cent and 18 per cent for $V\!=\!8$ and $V\!=\!9$, respectively.  

A clear degradation of the precision associated with the estimators is seen when considering the set of fainter objects ($V\!=\!9$). This is due to the fact that $V$ is directly coupled to the photon shot noise level and hence to $S/N$. 

Notice that no results have been plotted concerning the set characterised by $V\!=\!10$ as no sensible output was generated by the ACPS-modelling procedure. We should mention that at this $S/N$, the automated pipeline is no longer capable of supplying an input value for ${\rm{d}}\Delta\nu/{\rm{d}}n$ and hence the model p-mode spectrum generated by the ACPS-modelling procedure assumes a constant large separation with $n$.


\subsection{On retrieving $i$ and $\nu_{\rm{s}}$}
\begin{figure}
     \resizebox{\hsize}{!}{\includegraphics{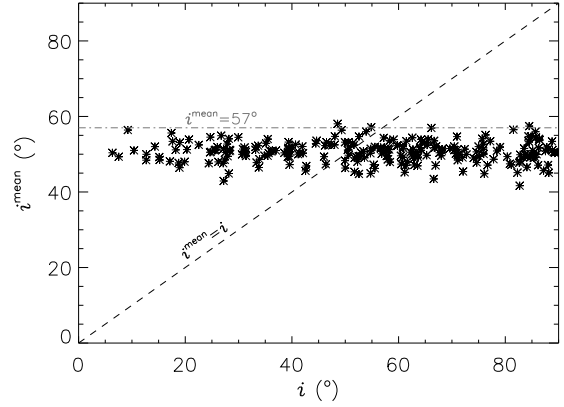}} 
     \caption{The posterior mean of the stellar inclination angle, $i^{\rm{mean}}$, versus the input inclination, $i$. This plot refers to the set of simulations with $V\!=\!8$.}
     \label{fig_inc}
 \end{figure}

With the current experimental setup, we face a scenario where the linewidths of the individual modes present in the simulated power spectra are systematically larger than the rotational splitting, i.e.~$\nu_{\rm{s}} \lesssim \Gamma$. Consequently, multiplet components are blended together, which strongly correlates the inclination with the rotational splitting, making a successful retrieval of these global parameters not feasible. As a further matter, the spectral resolution is too low ($\delta\nu\!=\!0.39\:{\rm{\mu Hz}}$) and will only marginally allow the rotational splitting to be resolved. By again looking at the PDF of the inclination for a single realisation of the power spectrum in Fig.~\ref{fig_Diagnose3}, we realise that we are in fact retrieving the imposed prior since there is no enough evidence in the data to proceed otherwise. Fig.~\ref{fig_inc} corroborates this, if we bear in mind that $i^{\rm{mean}} \equiv \int_{0\,^{\circ}}^{90\,^{\circ}} i \, p(i|D,I) \, \mathrm{d}i \simeq \int_{0\,^{\circ}}^{90\,^{\circ}} i \, p(i|I) \, \mathrm{d}i = \int_{0\,^{\circ}}^{90\,^{\circ}} i \, \sin(i) \, \mathrm{d}i = 57\,^{\circ}$. Similarly, by looking at the PDF of the rotational splitting for a single realisation of the power spectrum in Fig.~\ref{fig_Diagnose3}, we notice that we are in fact retrieving the imposed uniform prior, although slightly biased toward low values of the splitting.    

\begin{figure}
     \resizebox{\hsize}{!}{\includegraphics{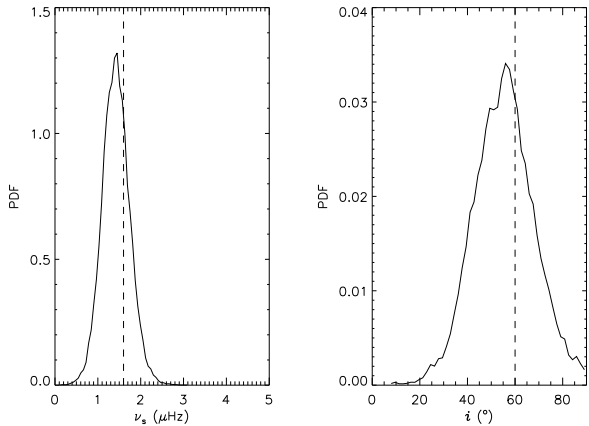}} 
     \caption{PDFs of $\nu_{\rm{s}}$ and $i$ resulting from the application of the ACPS-modelling procedure to a single realisation of the PDS of Boris. Reference values are given by $\nu_{\rm{s}}^{\rm{ref}}\!=\!4\,\nu_{{\rm{s}}\odot}\!=\!1.6 \: {\rm{\mu Hz}}$ and $i^{\rm{ref}}\!=\!60\,^{\circ}$, and are represented by dashed vertical lines. MAP estimates are in turn given by $\nu_{\rm{s}}^{\rm{MAP}}\!=\!1.44^{+ 0.26}_{- 0.36} \: {\rm{\mu Hz}}$ and $i^{\rm{MAP}}\!=\!55.9\,{^{\circ}}^{+ 11.8\,^{\circ}}_{- 13.2\,^{\circ}}$, therefore being within uncertainties of the reference values.}
     \label{fig_new}
 \end{figure}

In order to demonstrate the full capability of this tool, we have applied it to a series of higher $S/N$ realisations of the PDS of Boris (assumed to have $V\!=\!8$), which now correspond to 6-month long time series ($\delta\nu\!=\!0.06\:{\rm{\mu Hz}}$). Since we wish to avoid a scenario where $\nu_{\rm{s}} \lesssim \Gamma$ and enter a more favourable regime where instead $\Gamma < \nu_{\rm{s}} \lesssim \delta\nu_{02}$, we (i) increased the reference value for the mean rotational splitting, $\nu_{\rm{s}}^{\rm{ref}} \in \{2\,\nu_{{\rm{s}}\odot},4\,\nu_{{\rm{s}}\odot},6\,\nu_{{\rm{s}}\odot}\}$, while (ii) simultaneously adopting a narrower frequency interval of interest, ranging from 2100 to $3300 \: {\rm{\mu Hz}}$, at the low-frequency end of the acoustic spectrum. This latter step will effectively decrease the reference value for the mean linewidth, $\langle\Gamma\rangle^{\rm{ref}}$, since we use solar mode linewidths when generating the artificial power density spectra, which are known to increase toward higher frequencies. Finally, we adopted $i^{\rm{ref}}\!=\!60\,^{\circ}$ and imposed uniform priors on all the model parameters. Fig.~\ref{fig_new} illustrates a case where the inclination and the rotational splitting have been successfully retrieved, after applying the ACPS-modelling procedure to a single realisation of the PDS of Boris.

\section{Summary and Discussion}\label{sec:conclusions}
\begin{enumerate}

\item We have developed a new data analysis tool based on modelling and fitting the autocovariance of the acoustic PDS of a solar-type oscillator. Its main advantage when compared to a canonical peak-bagging procedure, relies on the fact that there is no need to carry out mode (angular degree $\ell$) identification prior to performing the fit. This procedure is in principle also suitable for being employed in the case of evolved stars displaying solar-like oscillations.

\item The implementation of the ACPS-modelling procedure has been thoroughly described. Furthermore, its automated character makes this procedure appropriate for the analysis of a large number of data sets (e.g.~arising from the \emph{Kepler} mission).    

\item The ACPS conveys information on the large and small frequency separations, mode lifetime and rotation. The current version of the ACPS-modelling procedure accepts $D_0$, $\nu_{\rm{s}}$, $\langle\Gamma\rangle$ and $i$ as free parameters. The prospective inclusion of $\Delta\nu$ and ${\rm{d}}\Delta\nu/{\rm{d}}n$ as additional free parameters is envisaged.

\item The tool has been applied to simulated data mimicking what would have been expected for a solar analog observed at high cadence during \emph{Kepler}'s survey phase. We assessed the potential biases as well as the precision associated with the several Bayesian summary statistics of the fitted parameters $D_0$ and $\langle\Gamma\rangle$, having been able to decide on a preferred summary statistic, namely, the MAP. We addressed a way of overcoming the inherent biases. These biases would not in any case underpin the usefulness of this procedure, since the preferred summary statistic is given together with robust estimates of the uncertainties: a MCMC sampler is employed in order to obtain the marginal posteriors for each of the model parameters, from which we can estimate the uncertainties. For instance, using the {\sc{seek}} routine (Quirion et al., in preparation) in order to estimate stellar parameters and basing it on a set of asteroseismic parameters returned by this procedure, would lead to the estimation of a set of stellar parameters possessing robust uncertainties \citep[see discussion in][]{Karoff10}. 

\item No sensible output has been generated by the ACPS-modelling procedure for the set with $V\!=\!10$. We could argue, of course, that by increasing the effective observational length of the simulated data, the method could successfully reach a lower $S/N$ since realisation noise is expected to scale as $1/\sqrt{T}$. 

\item Furthermore, retrieval of both $i$ and $\nu_{\rm{s}}$ could not be achieved with the current experimental setup and a reason for that was given based on the intrinsic characteristics of the simulated spectra. An example has however been included whereby the full capability of this tool is demonstrated. 

\end{enumerate}

\section*{Acknowledgments}
TLC is supported by grant with reference number SFRH/BD/36240/2007 from FCT/MCTES, Portugal. CK acknowledges financial support from the Danish Natural Sciences Research Council. WJC, YPE and SH acknowledge the support of the Science and Technology Facilities Council (STFC), UK. We wish to thank the anonymous referee for valuable comments.

\bibliographystyle{mn2e}
\bibliography{bibliography}

\appendix
\section{Parallel Tempering Metropolis--Hastings}\label{appendix}

\begin{figure}
	\begin{framed}
		\begin{algorithmic}[1]
		\small
		\Procedure{PT Metropolis--Hastings}{}
			\State $\bmath\Theta_{0,i} = \bmath\Theta_0 \, , \; 1 \leq i \leq n_\beta$
			\For{$t = 0,1,\ldots,n_{\rm{it}}-1$}
				\For{$i = 1,2,\ldots,n_\beta$}
					\State Propose a new sample to be drawn from a 
					\Statex \hspace{5em} proposal distribution: $\bmath\Lambda\sim N(\bmath\Theta_{t,i};\mathbfss{C}_i)$
					\State Compute the \emph{Metropolis ratio}: 
					\Statex \hspace{5em} $r=\frac{p(\bmath\Lambda|D,\beta_i,I)}{p(\bmath\Theta_{t,i}|D,\beta_i,I)} \frac{q(\bmath\Theta_{t,i}|\bmath\Lambda)}{q(\bmath\Lambda|\bmath\Theta_{t,i})}$
					\State Sample a uniform random variable: 
					\Statex \hspace{5em} $U_1\sim{\rm{Uniform}}(0,1)$
					\If{$r \geq U_1$}
						\State $\bmath\Theta_{t+1,i} = \bmath\Lambda$
					\Else
						\State $\bmath\Theta_{t+1,i} = \bmath\Theta_{t,i}$
					\EndIf
				\EndFor
				\State $U_2 \sim{\rm{Uniform}}(0,1)$
				\If{$1/n_{\rm{swap}} \geq U_2$}
					\State Select random chain: 
					\Statex \hspace{5em} $i\sim {\rm{UniformInt}}(1,n_\beta-1)$
					\State $j = i+1$
					\State Compute $r_{\rm{swap}}$:
					\Statex \hspace{5em} $r_{\rm{swap}}=\frac{p(\bmath\Theta_{t,j}|D,\beta_i,I) p(\bmath\Theta_{t,i}|D,\beta_j,I)}{p(\bmath\Theta_{t,i}|D,\beta_i,I) p(\bmath\Theta_{t,j}|D,\beta_j,I)}$
					\State $U_3 \sim{\rm{Uniform}}(0,1)$
					\If{$r_{\rm{swap}} \geq U_3$}
						\State Swap parameter states of chains $i$ and $j$:
						\Statex \hspace{7em} $\bmath\Theta_{t,i} \leftrightarrow \bmath\Theta_{t,j}$
					\EndIf
				\EndIf
			\EndFor
			\State \Return $\bmath\Theta_{t,i} \, , \; \forall t \, , \; i\!:\!\beta_i\!=\!1$ 
		\EndProcedure
		\end{algorithmic}
	\end{framed}
	\caption{Pseudocode-written version of the parallel tempering Metropolis--Hastings algorithm. $n_\beta$ is the total number of tempered chains, $n_{\rm{it}}$ is the number of iterations when running the MCMC, $q(\bmath\Lambda|\bmath\Theta_{t,i})\!=\!N(\bmath\Theta_{t,i};\mathbfss{C}_i)$ is a multivariate Gaussian distribution centred on $\bmath\Theta_{t,i}$ and with diagonal covariance matrix $\mathbfss{C}_i$, and $n_{\rm{swap}}$ is the mean number of iterations between successive proposals to swap the parameter states of two adjacent chains.}
	\label{fig_Pseudocode}
\end{figure}

\label{lastpage}

\end{document}